\documentclass[12pt,reqno]{article}
\usepackage{amsmath,epsfig,amsmath,amssymb,verbatim,mathrsfs,array,layout,textcomp,amssymb,latexsym}
\usepackage{bbm}
\usepackage[hypertex]{hyperref}
\usepackage{array,layout}
\usepackage{color}
\include{epstopdf}

\allowdisplaybreaks
\numberwithin{equation}{section}

\def\OMIT#1{{}}
\newcommand{\lsim}{\mathrel{\rlap{\lower4pt\hbox{\hskip1pt$\sim$}}
    \raise1pt\hbox{$<$}}}         
\newcommand{\gsim}{\mathrel{\rlap{\lower4pt\hbox{\hskip1pt$\sim$}}
    \raise1pt\hbox{$>$}}}         

\newcommand{\be}{\begin{equation}}
\newcommand{\ee}{\end{equation}}
\newcommand{\beq}{\begin{equation}}
\newcommand{\eeq}{\end{equation}}
\newcommand{\bea}{\begin{eqnarray}}
\newcommand{\eea}{\end{eqnarray}}

\begin{document}
\begin{titlepage}
\begin{center}
\hfill{\tt WIS/16/09-DEC-DPPA}\\
\vskip 20mm

{\Large{\bf Inverted Sparticle Hierarchies from \\
\vskip 5mm Natural Particle Hierarchies}}

\vskip 10mm

{\bf O.~Aharony, L.~Berdichevsky, M.~Berkooz,\\
Y.~Hochberg and D.~Robles-Llana\footnote{Emails: ofer.aharony, leon.berdichevsky, micha.berkooz, yonit.hochberg, daniel.robles@weizmann.ac.il}
}

\vskip 4mm
{\em Department of Particle Physics and Astrophysics,}\\
{\em Weizmann Institute of Science,}\\
{\em Rehovot 76100, Israel}\\
[2mm]

\end{center}
\vskip 2cm

\begin{center} {\bf ABSTRACT }\end{center}
\begin{quotation}
\noindent A possible resolution of the flavor puzzle is that the
fermion mass hierarchy can be dynamically generated through the
coupling of the first two generation fields to a strongly coupled
sector, which is approximately conformally invariant and leads to
large anomalous dimensions for the first two generation fields over
a large range of energies. We investigate the possibility of using
the same sector to also break supersymmetry. We show that this
automatically gives an ``inverted hierarchy'' in which the first two
generation squarks and sleptons are much heavier than the other
superpartners. Implementing this construction generically requires
some fine-tuning in order to satisfy the constraints on
flavor-changing neutral currents at the same time as solving the
hierarchy problem. We show that this fine-tuning can be reduced to
be milder than the percent level by making some technically natural
assumptions about the form of the strongly coupled sector and its
couplings to the standard model.

\end{quotation}

\vfill

\end{titlepage}

\eject

\section{Introduction}\label{sec:intro}
The standard model (SM), when embedded into a theory valid at energy
scales higher than a TeV, contains two types of unnaturally small
couplings: the mass-squared of the Higgs field (the hierarchy
problem), and many of the Yukawa couplings and mixing angles in both
the quark and lepton sectors (the flavor puzzle). Both issues can
be addressed in various ways, and some of the most elegant solutions
involve adding to the theory additional sectors that decouple at some
high energy scale.

A natural solution to the hierarchy problem is supersymmetry (SUSY),
which to avoid fine-tuning must be broken softly at energies not
much higher than the weak scale (see \cite{Martin:1997ns} for a
review). This breaking cannot happen within the simplest supersymmetric
generalization of the SM itself, and so an additional sector
breaking supersymmetry is usually required, as well as some mechanism for coupling
the two sectors such that SUSY breaking is transmitted to the SM.

The flavor puzzle can be addressed by introducing approximate
horizontal flavor symmetries which suppress the small Yukawa
couplings \cite{Froggatt:1978nt,Horizontal,Horizontalbis}. Another
way to solve the flavor puzzle is to consider a strongly coupled
and approximately conformal sector
\cite{Georgi:1983mq,Nelson:2000sn,Nelson:2001mq,Poland:2009yb}. The
quarks and leptons of the first two generations can obtain large
anomalous dimensions through direct couplings to this ``conformal
field theory (CFT) sector''.\footnote{We will refer to the sector
responsible for generating the flavor hierarchy as the CFT sector,
even though its dynamics is only approximately conformal in some
range of energies.} If the conformal regime persists for a large
range of energy scales, renormalization group (RG) evolution can
lead to realistic hierarchical Yukawa couplings at low energies,
starting from an unstructured (anarchical) UV theory. The resulting
Yukawa structures are similar to those obtained using horizontal
symmetries.

In principle the above solutions can be combined by making the
``flavor CFT sector'' supersymmetric, and some implications of this
were discussed in \cite{Nelson:2001mq,Kobayashi:2001kz}.\footnote{As discussed in
\cite{Nelson:2001mq,Kobayashi:2001kz}, if one assumes that the SUSY breaking is
transmitted to the flavor CFT sector above the scale where this
sector is superconformal, the RG flow of the CFT sector naturally
suppresses also SUSY-breaking flavor-violating terms, so it
helps in solving the supersymmetric flavor problem.} However, it is
natural to wonder whether it may be possible to use the flavor CFT
sector also for supersymmetry breaking. This could allow for the construction of
more economical models with fewer additional sectors that have not
yet been observed. Traditionally models with less unobserved sectors
have been preferred, following Occam's razor, though in the context
of a complete theory of nature like string theory it is not clear if
generic vacua, agreeing with all observations made so far, contain a small or a large
number of sectors. In this paper we will simply assume
that the same sector plays both roles described above --
it generates the flavor hierarchy through large anomalous dimensions
and it also breaks supersymmetry -- and we will analyze the
phenomenological implications of this assumption.

Our assumption, however, cannot be taken at face-value since it
immediately leads to a severe tension between an appropriate
suppression of flavor changing neutral currents (FCNCs) and
naturalness of the Higgs sector. On the one hand, the dynamical
scale of our CFT sector determines the scale of SUSY breaking,
and this cannot be too large without introducing fine-tuning of the
Higgs mass (for instance, the supersymmetric partner of the top
quark should not be much heavier than a TeV or so
\cite{Barbieri:1987fn,Dimopoulos:1995mi,Wright:1998mk}). On the other hand, and
unlike models with flavor-universal SUSY breaking, in our models the
first two generations must directly couple to the CFT sector in a
flavor non-universal way in order to generate the flavor hierarchy.
Generic non-universal flavor dynamics leads to large FCNCs unless it
decouples from the standard model fields at a rather heavy scale;
the experimental constraints on FCNCs place lower bounds on the
scale of flavor-violating new physics that are of the order of tens
of thousands of TeVs \cite{Bona:2007vi,Ciuchini:2007cw}.
In general, an even more serious problem would arise from the
experimental bounds on CP violation
\cite{Bona:2007vi,Ciuchini:2007cw}. In our analysis in most of this paper we consider the bounds from CP-conserving processes; in the final section we discuss the implications of the stronger bounds from CP violation.

The above fine-tuning problem does not have to be faced in its most
severe form, since in our models there is always some built-in
hierarchy between the scale of the stop mass and the scale of
flavor-violating effects involving the first two generations, to
which the strongest FCNC bounds apply. The superpartners of the
first two generations couple directly to the CFT sector and obtain
masses of the order of the SUSY breaking scale. The existence of
these direct couplings implies that the CFT sector must have a
symmetry group (which is a global symmetry group from the point of
view of the CFT sector) that contains the SM gauge group as a
subgroup. As we do not allow direct couplings for them (which would
destroy the flavor hierarchy), the third generation scalars,
gauginos and Higgs sector fields will obtain SUSY breaking masses
mostly through gauge-mediated couplings to the CFT sector
\cite{Giudice:1998bp,Meade:2008wd}. These masses will then be
suppressed by an extra small factor from SM gauge couplings. Such
inverted hierarchy models \cite{
Dine:1990jd,Pomarol:1995xc,Cohen:1996vb}, where the first two
generation scalars are relatively heavy, have been considered before
as a possible resolution of the supersymmetric flavor problem, from
a variety of different perspectives
\cite{Invertedanomalous,Invertedanomalous2,Flavorsusy,
radiativeinverted,Compositemodels,miscellaneousinverted,Franco:2009wf,Craig:2009hf}.
In our models this feature is an automatic outcome of our
assumptions.

Unfortunately, the natural inverted hierarchy obtained from the above
considerations is still not large enough to reduce the fine-tuning
to acceptable levels. One reason for this is that our CFT sector
generally contains many fields charged under the SM gauge group, so
that in the language of gauge mediation we have a large effective
number of messengers \cite{Martin:1996zb}, reducing the
suppression of the gauge-mediated superpartner masses compared to
the directly-mediated masses. Another issue to address is a generic problem in inverted hierarchy models, where the two-loop RG
flow often leads to negative masses-squared for some
of the third generation sfermions, which must be avoided
\cite{ArkaniHamed:1997ab,Agashe:1998zz,Hisano:2000wy}.

We are thus led to make several additional assumptions. We first
require that the CFT sector does not directly couple the first
generation and the second generation fields (``separable CFT
sectors''). This can be accomplished either by having a
``decomposable CFT sector", in which the fields coupled to the first
generation do not couple to the second and vice versa (implying a
factorization of the CFT sector gauge group), or by introducing some
horizontal-type symmetry which suppresses couplings between the
first two generations. The assumption of separable CFT sectors plays
two important roles:

\begin{itemize}
\item{} It helps to suppress FCNCs. Naively FCNCs are completely absent in such a setup,
but one must keep in mind that the decoupling between the
generations discussed here is in the interaction basis, so in the
mass basis some flavor-non-diagonal couplings suppressed by powers
of the mixing angles between the first two generations are always
present. (For quarks, these are typically of order the Cabibbo angle
$\sim 0.22$ or smaller.)  Thus, separability helps alleviate the
fine-tuning problem, but does not solve it completely.

\item{} It helps to solve the ``graceful exit'' problem in models of
dynamically-generated flavor hierarchies as in \cite{Nelson:2000sn,
Nelson:2001mq}. It was emphasized there that the dynamics
responsible for the breaking of the conformal symmetry and the
superconformal $U(1)_R$ symmetry
of the CFT sector  should not generate a large mixing in the
kinetic terms of the first two generation fields, since this would
wash out the hierarchy generated between their Yukawa couplings in
the conformal range of energies. The assumption of separable CFT
sectors automatically solves this problem, as it implies that there
is no additional source of mixing between the first two generation
fields.
\end{itemize}

Even if the CFT sector is separable, we are still left with a rather
large amount of fine-tuning, so we then consider two additional ways
to reduce this in our models.  First, we assume the scale of SUSY
breaking $\sqrt{F}$ to be much smaller than the scale $M$ at which
the SM fields decouple from the CFT sector. This hierarchy can be
achieved -- as in traditional models of gauge mediation
\cite{Dine:1981gu,Dine:1993yw} -- if the fields in the CFT sector
coupling to the standard model fields have a mass which is larger
than that of other fields in the CFT sector involved in the SUSY
breaking, or by some other mechanism which suppresses the SUSY
breaking scale \cite{Witten:1981nf}. This allows us to raise the
scale $M$ at which generic flavor-violating operators are generated,
though the fact that in these models the first two generation
sfermions are much lighter than the scale $M$ implies that one must
separately worry about flavor-violating effects involving sfermion
loops \cite{Hagelin:1992tc,Gabbiani:1996hi}. Second, we assume that
only some of the first two generation fields directly couple to the
CFT sector, while others do not; in particular we consider models
where only the left-handed superfields couple directly to the CFT
sector, or only the right-handed superfields, or only the
superfields which are in the $\bf 10$ representation of $SU(5)$ (in
the language of $SU(5)$ grand unified (GUT) models). Such partial couplings
are natural in brane realizations of the standard model. We are then
able to suppress the strongest flavor-violating effects coming from
chirality mixing operators in the down sector, and still solve the
flavor puzzle by generating large anomalous dimensions for those
fields which do couple directly to the CFT sector.

After making all the assumptions mentioned above, we find that only
a small fine-tuning (milder than the percent level) is necessary in
order to avoid FCNCs and solve the hierarchy problem, taking all
couplings and correlation functions of the CFT sector to be
otherwise generic. While the CFT sector must obey many requirements,
all of these are technically natural and could arise in the context
of some complete theory of high-energy physics.
Finding an explicit model satisfying all these conditions is
complicated, and we do not consider it here, but there is no
apparent obstruction to such constructions.\footnote{See
\cite{Ibe:2005pj} for recent examples of superconformal models with
dynamical SUSY breaking, and \cite{Schmaltz:2006qs} in the context
of metastable SUSY breaking.}

The outline of this paper is the following.  We begin in Section
\ref{sec:rev} by reviewing how flavor hierarchies can be generated
by a CFT, as well as the experimental constraints on FCNCs. In
Section \ref{sec:models} we present a general discussion of our
models in which all of the first two generation fields are coupled
to the CFT sector, and we introduce the notion of a separable CFT
sector. We briefly consider one-scale models, $F \sim M^2$, and then
move to a more detailed analysis of two-scale models, $F \ll M^2$.
In Section \ref{sec:1c} we discuss partially coupled scenarios,
which result in less fine-tuning, and present some examples of the
spectra of superpartners arising from models of this type. We end in
Section \ref{sec:conc} with a summary of our results and
conclusions.

\section{Review}\label{sec:rev}

In this section we review the two main tools of our constructions.
Section \ref{ssec:revNS} reviews the Nelson-Strassler mechanism,
namely how a CFT sector can explain the intergenerational mass
hierarchy.  In Section \ref{ssec:revFCNC} we outline the main
experimental FCNC bounds derived from general dimension-six
flavor-violating operators and from diagrams with squark-gluino
boxes. These will be basic constraints on all of our models.

\subsection{CFTs and flavor hierarchies}\label{ssec:revNS}

\subsubsection{The Nelson-Strassler Mechanism}

In \cite{Nelson:2000sn}, Nelson and Strassler considered a mechanism
that dynamically generates the IR hierarchical structure of the
minimal supersymmetric standard model (MSSM) Yukawa couplings
starting from a flavor anarchical UV theory.\footnote{We assume for
simplicity that the SUSY generalization of the standard model
contains only the MSSM. Our considerations are easily generalized
when additional fields are present.} In these models, short-distance
Yukawa couplings are arbitrary order one matrices, and the pattern
of masses and mixing angles at low energies arises as the result of
renormalization group flow.

The basic idea is the following. Unlike in asymptotically free
theories in the weak coupling regime, where dimensionless couplings
have logarithmic running, in a nearly scale-invariant theory such
couplings can have power law running. Furthermore, if the theory is
not weakly coupled, the anomalous dimensions of such couplings can
be significant. Following an initial idea put forward in
\cite{Georgi:1983mq}, the Nelson-Strassler mechanism is the
application of this idea to the Yukawa couplings. When light quarks
and leptons have direct couplings to a strongly coupled conformal
sector, they can acquire substantial anomalous dimensions. As a
result, a large Yukawa hierarchy can be generated if the conformal
regime persists over a wide enough range of energy scales, and if
the anomalous dimensions of these fields are appreciable.

We will assume below that the fixed point theory, near which the RG
flow occurs, has ${\cal N}=1$ SUSY. In this case the anomalous
dimensions, and hence the IR Yukawa couplings, are determined by the
$U(1)_R$ charges of the fixed point theory \cite{Flato:1983te}. This
$U(1)_R$ is approximate and may be accidental from the UV point of
view. The resulting masses and mixing angles resemble those
predicted in models with horizontal Abelian symmetries, but they
differ from these models in that the R-symmetry which sets the
charges arises dynamically and accidentally. Furthermore, the
characteristic ratio between different Yukawas is not set by a vacuum
expectation value (VEV)
of a field, but by the number of energy decades spent near the fixed
point.

To be more concrete and to set the notation, we consider an
${\mathcal N}=1$ supersymmetric theory with gauge group $\tilde G$,
charged matter $Q$, neutral matter $X$ and a superpotential
$W(X,Q)$, and we will assume that it flows to a conformal theory in
the infrared. Here $Q$ is not to be confused with the SM quarks,
which are CFT sector singlets and are included in the $X$ fields.

By unitarity all gauge-invariant operators (except the identity
operator) have scaling dimensions ${\rm dim}[{\mathcal O}(Q,X)]\geq 1$
\cite{Mack:1975je}, with a strict inequality when ${\mathcal O}$ is
not a free field. As a consequence, $Q$ may have a negative
anomalous dimension but $X$, being gauge-invariant, always has
anomalous dimension $\gamma_X=2[{\rm dim}[X]-1]\geq 0$. Consider now
superpotential terms of the form $c_{k,\mathcal{O}} X^k {\mathcal
O}(Q)$, where ${\mathcal O}(Q)$ is a non-trivial $\tilde
G$-invariant operator built from the charged fields $Q$, which
therefore has ${\rm dim}[{\mathcal O}(Q)] > 1$. At or near the
superconformal fixed point, most of these terms are irrelevant since
${\rm dim}[X^k{\mathcal O}(Q)]>k+1$, and thus if $k \geq 2$ the
coupling $c_{k,\mathcal{O}}$ will flow to zero. Conversely, relevant
superpotential terms can drive the theory to a fixed point in which
fields of type $X$ possess positive anomalous dimensions. We will be
interested in terms of the form $X{\mathcal O}(Q)$ (for ${\rm
dim}[{\mathcal O}]<2$), which can exist at a non-trivial fixed point
when $0<\gamma_X<2$.

If the superconformal sector has a global symmetry $G$ -- in which
the standard model gauge group $G_{SM}\equiv SU(3)_c\times
SU(2)_L\times U(1)_Y$ can be embedded -- one can consider the
following scenario. The effective theory has gauge group $\tilde
G\times \hat G$, with $\tilde G$ strongly coupled and an appropriate
subgroup $\hat G$ of $G$, which contains $G_{SM}$, weakly gauged.
As far as ${\tilde G}$ is concerned the SM quarks and leptons
$L^i, R^j$ $(i,j=1,2)$ of the first two generations are of type
$X$ above -- they are charged under $G$ but neutral under $\tilde G$, and they
can directly couple in the superpotential to charged matter of $\tilde G$ by couplings of the form $W = X {\mathcal O}$.
In this case, since they are $X$-type fields, the conformal dynamics associated to
$\tilde G$ may cause them to develop large positive anomalous dimensions.
These will in turn cause their standard model Yukawa couplings
$W = y_{ij} L^i H R^j$ (where we canonically normalize the fields) to run as\footnote{This is the running during the range of energies where the CFT
sector is conformal. In addition, \eqref{yuksup} will obtain order one corrections
from the entrance to and exit from this regime.}
\be \label{yuksup}
y_{ij}(\mu) =
y_{ij}(\mu_0)\left(\frac{\mu}{\mu_0}\right)^{\frac{1}{2}[\gamma(L^i)
+ \gamma (R^j)]}. \ee

We can now suppose that all Yukawa couplings are unstructured and of
order one at some high scale, say the Planck or string scale $M_0$. If the gauge
couplings of $\tilde G$ and the direct couplings become conformal
near some scale
$\mu_0=M_>$ and remain so until some lower scale $\mu = M_<$, with $\epsilon \equiv M_< / M_> \ll 1$, then the standard
model Yukawa couplings will run down to
\be\label{yukyuk} y_{ij}(M_<)\sim
\left(\frac{M_<}{M_>}\right)^{\frac{1}{2}(\gamma_{Li}+\gamma_{Rj})}
= \epsilon^{\frac{1}{2}(\gamma_{Li}+\gamma_{Rj})}\equiv
\epsilon_{L_i}\epsilon_{R_j}. \ee
We assume here (this will later be slightly modified) that the CFT
sector has a mass gap of order $M_<$, so that below this scale we
are left only with the weakly coupled MSSM fields, with no remaining
mixing to the CFT sector.

The above factorized texture of Yukawa couplings, with each fermion
coming with its own suppression factor, leads to the following
structure of mixing angles
\cite{Froggatt:1978nt,Horizontal,Horizontalbis,Nelson:2000sn}
\be (V^q_{L})_{ij} \sim \frac{\epsilon_{L_i}}{\epsilon_{L_j}}\sim
(V^q_{L})_{ji}\,,~~~ (V^q_{R})_{ij} \sim
\frac{\epsilon_{R_i}}{\epsilon_{R_j}}\sim
(V^q_{R})_{ji}~~~(i<j)\,,~~~q=u,d,\ell\,,
\ee
and fermion mass relations
\be \frac{m_i}{m_j}\sim
\frac{\epsilon_{L_i}\epsilon_{R_i}}{\epsilon_{L_j}\epsilon_{R_j}}\,,
\ee
where we assume without loss of generality $\gamma_{L1} \geq \gamma_{L2}$,
$\gamma_{R1} \geq \gamma_{R2}$.
Combining the two above equations in the case of the quarks one obtains
\be\label{eq:NSdiagmat}
 (V^q_{L,R})_{ii}\sim1,\ \ \
(V^q_L)_{ij}\sim|V_{ij}|,\ \ \
(V^q_R)_{ij}\sim\frac{m_{q_i}/m_{q_j}}{|V_{ij}|}\ \ \ (i<j),
~~q=u,d, \ee
where $V_{ij}$ is the CKM matrix $V = V^u_L V^{d\dagger}_L$.

It is essential for our purposes that the measured Yukawa hierarchy
can then be explained by coupling only the first two generations to
the CFT. To this end, suppose a CFT has a spectrum of chiral
operators ${\mathcal O}_{1,2}$ with distinct scaling dimensions
$d_{1,2}$ related to their R-charges $r_{1,2}$ as
$d_{1,2}=\frac{3}{2}r_{1,2}$, and with $r_1 < r_2 < 4/3$. If some
linear combination of MSSM superfields couples to ${\mathcal O}_1$
and this coupling flows to a non-trivial fixed point, then at the
fixed point this linear combination (which we can call the first
generation and denote by $\Phi_1$) will have $r_{\Phi_1} = 2 - r_1$
and $d_{\Phi_1} = 3 - d_1$. If some other linear combination of MSSM
superfields couples to ${\mathcal O}_2$, then this other linear
combination will contain a component along the $\Phi_1$ direction in
flavor space whose coupling will be irrelevant at the fixed point,
and a component orthogonal to $\Phi_1$ which we can denote by
$\Phi_2$ (and which will be the second generation). Assuming that
the ${\mathcal O}_2 \Phi_2$ coupling also flows to a fixed
point\footnote{We assume that this new fixed point is stable, namely
that it does not have any relevant operators that can appear in the
K\"ahler potential. In our examples we will typically know that the
original CFT before coupling to the MSSM fields is stable, and it
seems reasonable to assume that this remains true after this
coupling, but we do not prove this.}, we will have at that fixed
point $r_{\Phi_2} = 2 - r_2$ and $d_{\Phi_2} = 3 - d_2$. By
construction there are no R-symmetry violating operators of the form
${\mathcal O}_1 \Phi_2$. The R-symmetry violating operators
${\mathcal O}_2\Phi_1$ are irrelevant, so after the RG flow they
will be suppressed by factors of order
$\epsilon^{\frac{1}{2}(\gamma(\Phi_1)-\gamma(\Phi_2))}$. One can
then start with a generic gauge-invariant K\"ahler potential and
superpotential in the UV, and the conformal dynamics will guarantee
that in the effective low-energy theory at the scale $M_<$ the
K\"ahler potential and superpotential are approximately flavor
diagonal in the CFT basis \cite{Nelson:2000sn}, and exhibit a
hierarchy in the Yukawa couplings.

\subsubsection{Some further assumptions and requirements}

After the conformal dynamics has done its job, some relevant
deformation must drive the theory away from the fixed point. For
example, the escape from the conformal window can be the result of
small masses for some CFT sector particles, or of confinement of the
CFT sector gauge group. It is important, however, that threshold
effects at $M_<$ do not spoil the hierarchy of the Yukawa couplings
built during the conformal regime. In particular, flavor
off-diagonal wavefunction renormalizations ${\mathcal Z_{ij}}$
should be small. This is the concept of a graceful exit \cite{Nelson:2000sn,Nelson:2001mq}. In our constructions this will be guaranteed
through the notion of separable CFT
sectors. In such scenarios, described in Section \ref{ssec:sephid} below,
intergenerational couplings are absent in
the interaction (CFT) basis, up to terms similar to the previously
discussed ${\mathcal O}_2\Phi_1$. Such a term can induce off-diagonal
wavefunction renormalizations of order $\epsilon_1/\epsilon_2$ (where
$\epsilon_1$ and $\epsilon_2$ are typical suppression factors for the
operators of the first two generations), which are small enough such that the Yukawa hierarchy in the canonical basis is not offset.
Note that this implies
that in our models we expect to have flavor violating terms at least of
order $\epsilon_1/\epsilon_2$.

We should also worry about proton decay. As emphasized in
\cite{Nelson:2000sn,Nelson:2001mq}, if the CFT sector contains couplings
violating baryon number, the graceful exit must take place above a
scale $M_<\sim 10^{15}\,{\rm GeV}$ in order to appropriately
suppress proton decay from dimension six operators.  Dimension five
operators pose additional constraints. It is argued there, that even
assuming that these terms are not generated by the CFT, they can
already be present at $M_{pl}$ (R-parity can be imposed to forbid
dimension four couplings, but it does not preclude the presence of
dimension five operators). In that case the natural suppression
factor $1/M_{pl}$ has to be strengthened by small coefficients in
the range $(10^{-6}-10^{-7})$ \cite{Nelson:2000sn,Hisano:1992jj}. In
many of the scenarios considered by Nelson-Strassler (at small ${\rm
tan}\beta$) this is achieved by the same suppression factors
$\epsilon_i$ in \eqref{yukyuk} responsible for the Yukawa hierarchy.
We will assume in our constructions that the CFT sector preserves baryon
number and that the appropriate suppression of dimension five operators at
$M_{pl}$ proceeds along the same lines as just discussed.  This allows for low $M_<$ and no proton decay.

To illustrate part of the above discussion, we now review two
examples presented in \cite{Nelson:2000sn}. We note that none of
these models fully satisfies the requirements discussed above that
need to be imposed on our CFT sectors. Apart from the fact that they
do not break SUSY, the first example is separable but
baryon-violating, whereas the second respects baryon number but is
not separable. In this paper we are interested in a qualitative
discussion on general classes of possible models, and are not
concerned with concrete model building. We assume that a model
satisfying all our requirements can be built.

\subsubsection{An {$SU(3)^3$ example}}\label{SU3}

We first present an example which is separable but can induce proton
decay.  As mentioned, the scale $M_<$ must then be above $\sim
10^{15}$ GeV in order to suppress baryon number violating dimension
six operators in the effective K\"ahler potential.

In this model the full symmetry group of the CFT sector is
$SU(3)^3\times {\bf Z}_3\times SU(5)\times
SU(4)$.  The group $SU(5)\times SU(4)$ is a
gauge symmetry of the CFT sector under which all MSSM fields are
neutral. These groups become strongly coupled at scales $\Lambda_5$
and $\Lambda_4$ respectively, and the couplings $g_5$ and $g_4$
eventually reach a fixed point. $SU(3)^3$ is a global symmetry of
the CFT sector. After weakly gauging an appropriate subgroup, the
first $SU(3)$ factor is identified with the color gauge group, the
electroweak $SU(2)_L$ group resides in the second $SU(3)$, and hypercharge is
embedded in the second and third $SU(3)$'s. $SU(3)^3$ can be seen as
a subgroup of the GUT group $E_6$ \cite{Gursey:1975ki}, and the MSSM quarks
and leptons are contained in three copies of $27\equiv(3,\bar
3,1)+(\bar 3,1,3)+(1,3,\bar 3)$. In addition to the (grand unified
generalization of the) usual chiral
superfield content of the MSSM, this example contains chiral
superfields $Q$ ($\bar Q$) and $Q'$ ($\bar Q'$) which transform in
the fundamental (anti-fundamental) representations of $SU(4)$ and
$SU(5)$, respectively. Including appropriate Higgs multiplets, the field content of
the model is summarized in Table \ref{tab1}.

\begin{table}[t]
\begin{center}
\begin{tabular}{|c|c|c|c|c|} \hline\hline
\rule{0pt}{1.2em}%
& $SU(3)^3$ & $SU(4)$ &  $SU(5)$ & {\rm dimension} \cr \hline\hline
$27_1,~27_2,~27_3$ & $(3,\bar 3,1)+(1,3,\bar 3)+(\bar 3,1,3)$ & $1$&
$1$ & $\frac{5}{3},\frac{4}{3},1$\cr $27_H,~27'_{H}$ & $(3,\bar
3,1)+(1,3,\bar 3)+(\bar 3,1,3)$ & $1$ & $1$ & $1,1$ \cr
$\bar{27}_H,~\bar{27}'_{H}$ & $(\bar3,3,1)+(1,\bar3, 3)+(
3,1,\bar3)$ & $1$& $1$ & $1,1$ \cr $Q$ & $(3,1,1)+(1,3,1)+(1,1,3)$ &
$4$ & $1$ & $\frac{5}{6}$ \cr $\bar Q$ &
$(\bar3,1,1)+(1,\bar3,1)+(1,1,\bar3)$ & $\bar4$ & $1$ &
$\frac{5}{6}$\cr $Q'$ & $(3,1,1)+(1,3,1)+(1,1,3)$ & $1$ & $5$ &
$\frac{2}{3}$\cr $\bar Q'$ & $(\bar3,1,1)+(1,\bar3,1)+(1,1,\bar3)$ &
$1$ & $\bar5$ & $\frac{2}{3}$ \cr \hline\hline
\end{tabular}
\end{center}
\caption{Quantum numbers and scaling dimensions at the fixed point
of the chiral superfields in the $SU(3)^3$ model.}\label{tab1}
\end{table}

Apart from gauge and Yukawa interactions this model contains the
following gauge and ${\bf Z}_3$-invariant superpotential:
\be\label{superpot} W=\sum_{i=1,2}\lambda_i \bar Q Q 27_i
+\lambda'\bar Q' Q' 27_1. \ee
Here we used flavor rotations between the three generations to
ensure that only the first generation couples to $\bar Q' Q'$ and
only the first two to $\bar Q Q$, as described above; the third
generation then does not couple directly to the CFT sector. With
this convention the MSSM fields acquire the anomalous dimensions
listed in Table \ref{tab1}, as we will now describe.

Going down in energies, the dynamics of this theory can be analyzed
in terms of scales $\Lambda_5$ and $\Lambda_4$ (we take
$\Lambda_5>\Lambda_4$). The $SU(5)$ group has nine charged flavors,
which drive the theory to a Seiberg conformal fixed point (since
$\frac{3}{2}N_c<N_f<3N_c$) \cite{Seiberg:1994pq}. At this fixed
point the dimensions of gauge-invariant chiral operators are related
to their R-charges as $d=\frac{3}{2}r$ through the superconformal
algebra of the fixed point theory. The R-charges can in turn be read
from the microscopic theory and result in $d_{\bar
Q'}=d_{Q'}=\frac{3}{2}\frac{N_f - N_c}{N_f}=\frac{2}{3}$. Therefore
the fields $\bar Q'$ and $Q'$ acquire negative anomalous dimensions
 of $\gamma_{\bar Q', Q'}=-2/3$ at the fixed point, making the coupling
 $\lambda'$ relevant and driving the theory to a new fixed point
 where the last term in the superpotential \eqref{superpot} is
 marginal. Assuming the existence of this fixed point, and given the anomalous dimensions
 of  $\bar Q'$, $Q'$, the field $27_1$ must acquire a positive
 anomalous dimension of $\gamma_{27_1}=4/3$. This in turn causes
 $\lambda_1$ and the coupling of the $27_1$ to the Higgs field to become
 irrelevant and highly suppressed at low energies.

As we continue flowing, the $SU(4)$ theory becomes strongly coupled at the scale $\Lambda_4$,
 and its dynamics is also superconformal below this energy scale. In
 a similar fashion to the primed fields, $\bar Q$ and $Q$ acquire
 negative anomalous dimensions of $\gamma_{\bar Q, Q}=-1/3$, causing the coupling
 $\lambda_2$ to become relevant, while $\lambda_1$ remains irrelevant. The coupling $\lambda_2$ drives the theory to
 a fixed point where the $\lambda_2$ term in the superpotential
 \eqref{superpot} is marginal. The anomalous dimension of the $27_2$
 field at the fixed point is $\gamma_{27_2}=2/3$, causing its Yukawa coupling to the
 Higgs fields to be suppressed at low energies.

 The theory can naturally exit the conformal regime, for example, through
 confinement of an additional strong group $H$ with no extra matter charged under
 it.
If we have non-renormalizable couplings, generated at some high scale $M_* \geq M_>$, of the form
\be \label{inducedmass}
 \int\,d^2\,\theta\,\frac{1}{M_*^2} {\rm tr}(W_{\alpha}^2)\,\left(\bar Q Q+\bar Q'
 Q'\right)
 \ee
where $W_{\alpha}$ are the gauge superfields of $H$, then this gives
 small masses to the $Q,\bar Q, Q',\bar Q'$
 fields after condensation of the $H$ gauginos contained in $W_{\alpha}^2$, and leads $SU(5)\times SU(4)$ to a confining
 phase.\footnote{The term \eqref{inducedmass} is slightly enhanced by
the negative anomalous
 dimension of the quark bilinears, but it can still easily give small masses.}

Since the anomalous dimension of $27_1$ is twice that of $27_2$,
this model makes the generic prediction that $\epsilon_1\sim
\epsilon_2^2$ (assuming for simplicity that $\Lambda_5\sim\Lambda_4$ and that both strong groups
exit the conformal window around the same scale). However, since
in this model suppression factors are universal within a generation,
it predicts $m_u/m_t\sim m_d/m_b$, a prediction that is off by around two orders of magnitude. For more details on this model, see \cite{Nelson:2000sn}.

\subsubsection{A ``10-centered'' example}\label{10centered}

In this model the CFT sector does not induce proton
decay and the  exit scale $M_<$ could in principle be as low as $10
\,{\rm TeV}$. The gauge groups and matter content are listed in
Table \ref{tab2}, classifying fields according to their representations
in $SU(5)$ GUTs.

\begin{table}[h]
\begin{center}
\begin{tabular}{|c|c|c|c|c|} \hline\hline
\rule{0pt}{1.2em}%
& $SU(5)_{GUT}$ & $Sp(8)$ &  $Sp(8)'$ & {\rm dimension} \cr
\hline\hline $T_{1,2,3}$ & $10$ & $1$& $1$ &
$\frac{42}{25},\frac{69}{50},1$\cr $\bar F_{1,2,3}, \bar H$ & $\bar
5$ & $1$ & $1$ & $1$ \cr $H$ & $5$ & $1$& $1$ & $1$ \cr $Q$ &
$\overline{10}$ & $8$ & $1$ & $\frac{87}{100}$ \cr $A$ & $1$ & $27$ & $1$
& $\frac{3}{5}$\cr $J, K, L, M$ & $1$ & $8$ & $1$ & $\frac{3}{4},
\frac{3}{4}, \frac{3}{4}, \frac{9}{20}$\cr $\bar Q'$ & $10$ & $1$ &
$8$ & ${\rm confined}$ \cr $R, S$ & $1$ & $1$ & $8$ & {\rm
confined}\cr \hline\hline
\end{tabular}
\end{center}
\caption{Quantum numbers and scaling dimensions of chiral
superfields in the {\bf 10}-centered model.}\label{tab2}
\end{table}

The usual MSSM superfields are $T_{1,2,3}, \bar F_{1,2,3}, \bar H, H$.
The gauge-invariant superpotential of the model is schematically given by
\be W=T_1 QL+ \sum_{i=1,2} T_i QM+A^5+(JK)(JK)+A^3LM+(MJ)(MK). \ee
Note that in this model only the standard model particles in the ${\bf 10}$
representation of $SU(5)$ couple directly to the CFT sector.
Relevant interactions such as $A^3$ and $A^4$ must be forbidden by
discrete symmetries, or be initially very small for some reason.
In this model the
$Sp(8)$ group flows to a fixed point where the above superpotential
is marginal, while the $Sp(8)'$ group confines at low energies.

From the anomalous dimensions of the fields $T_1$ and $T_2$ one can
see that this model makes the prediction
$\epsilon_{10_1}=\epsilon_{10_2}^{34/19}$ which is in quite good
agreement with predictions from $SU(5)$ models (see \cite{Nelson:2000sn}).

\subsection{Flavor Changing Neutral Currents}\label{ssec:revFCNC}
Generic models of new flavor physics introduce flavor violating
terms beyond the SM. Such terms are highly constrained by
experimental FCNC measurements. The most stringent constraints in
our models will arise from either the $K^0-\overline{K^0}$ system
\cite{Hagelin:1992tc,Gabbiani:1996hi,Bona:2007vi} or the
$D^0-\overline{D^0}$ system \cite{Gedalia:2009kh,DsystemExp}. Lepton
number violating processes will generically give much milder
constraints, so we confine this general discussion to $\Delta S=2$
and $\Delta C=2$ processes.

The effective Hamiltonians for $\Delta F=2$ processes at the scale
of new flavor physics $\Lambda_{NP}$ can be written as
\cite{Gabbiani:1996hi}
\be\begin{split}\label{NP}
 {\mathcal H}_{Eff}^{\Delta S=2}& =\frac{1}{\Lambda_{NP}^2}\left(\sum_{I=1}^5 z_I^K {\mathcal Q}^{sd}_I
  +\sum_{I=1}^3 \tilde z_I^K\tilde{\mathcal
  Q}^{sd}_I\right),\\
{\mathcal H}_{Eff}^{\Delta C=2}&
=\frac{1}{\Lambda_{NP}^2}\left(\sum_{I=1}^5 z_I^D{\mathcal Q}^{cu}_I
  +\sum_{I=1}^3 \tilde z_I^D\tilde{\mathcal
  Q}^{cu}_I\right),
\end{split}
\ee
where the 4-fermi operators are
\be
\begin{split}\label{eq:4ferops}
{\mathcal Q}_1^{q_iq_j} & =\bar q^\alpha_{L_j}\gamma_\mu
q^\alpha_{L_i}\bar q^\beta_{L_j}\gamma^\mu q_{L_i}^\beta, \qquad
{\mathcal Q}_2^{q_iq_j} = \bar q^\alpha_{R_j} q^\alpha_{L_i}\bar
q^\beta_{R_j} q_{L_i}^\beta, \qquad
{\mathcal Q}_3^{q_i q_j} = \bar q^\alpha_{R_j} q^\beta_{L_i}\bar q^\beta_{R_j} q_{L_i}^\alpha, \\
{\mathcal Q}_4^{q_iq_j} & = \bar q^\alpha_{R_j} q^\alpha_{L_i}\bar
q^\beta_{L_j} q_{R_i}^\beta, \qquad \qquad
{\mathcal Q}_5^{q_iq_j} = \bar q^\alpha_{R_j} q^\beta_{L_i}\bar q^\beta_{L_j} q_{R_i}^\alpha, \\
\end{split}
\ee
and where $\tilde {\mathcal Q}_{1,2,3}$ are obtained from ${\mathcal
Q}_{1,2,3}$ through $L\leftrightarrow R$. The dimensionless coefficients
$z_I^{K,D}, \tilde z_I^{K,D}$ encode the details of the new
physics generating the above operators at $\Lambda_{NP}$.

In order to compare to experiment, the above effective Hamiltonians
have to be evolved from the scale of new physics $\Lambda_{NP}$ down
to the $m_K$ and $m_D$ scales. Taking into account QCD running and
operator mixing one obtains (we do not write the tilded part of the
effective Hamiltonian, but it is implicit in our discussion)
\be\begin{split}\label{QCDH} \langle\overline{K^0}|{\mathcal H}^{{\Delta
S=2}}_{Eff}|K^0\rangle_I
&=\frac{1}{\Lambda_{NP}^2}\sum_{j=1}^5\sum_{r=1}^5\left(b_j^{(r,I)(K)}+\eta
c_j^{(r,I)(K)}\right)\eta^{a_j^{(K)}}z_I^{K} \langle\overline{
K^0}|{\mathcal
Q}^{sd}_r|K^0\rangle,\\
 \langle\overline{D^0}|{\mathcal H}_{Eff}^{\Delta C=2}|D^0\rangle_I
&=\frac{1}{\Lambda_{NP}^2}\sum_{j=1}^5\sum_{r=1}^5\left(b_j^{(r,I)(D)}+\eta
c_j^{(r,I)(D)}\right)\eta^{a_j^{(D)}}z_I^{D}\langle\overline{D^0}|{\mathcal
Q}^{cu}_r|D^0\rangle,\end{split}\ee
where $\eta\equiv \alpha_3(\Lambda_{NP})/\alpha_3(m_t)$, and the
other relevant inputs can be found in \cite{Ciuchini:1998ix} for the
$K^0-\overline{K^0}$ system and in \cite{Bona:2007vi} for the
$D^0-\overline{D^0}$ system. Throughout this paper we only use the bounds
coming from CP-conserving processes, see the discussion in the final section.
Imposing that the new physics contribution is not larger than the measured
values of $\Delta m_K=2\,{\rm Re}\langle K^0|{\mathcal
H}_{eff}^{\Delta S=2}|\overline{K^0}\rangle\simeq
3.483\cdot10^{-12}$ MeV and $\Delta m_D=2\,{\rm Re}\langle
D^0|{\mathcal H}_{eff}^{\Delta C=2}|\overline{D^0}\rangle\simeq
1.89\cdot10^{-11}$ MeV \cite{DsystemExp,Amsler:2008zzb}, the
following constraints are obtained:
\be
\begin{split}\label{eq:4ferbounds}
 \Lambda_{NP}&\gsim\sqrt{|z^K_1|}\, 1.0\times 10^3 \,{\rm TeV},\quad
 \Lambda_{NP}\gsim\sqrt{|z^K_2|} \,7.3\times 10^3 \,{\rm TeV},\quad
 \Lambda_{NP}\gsim\sqrt{|z^K_3|} \,4.1\times 10^3 \,{\rm TeV},\\
 \Lambda_{NP}&\gsim\sqrt{|z^K_4|} \,17 \times 10^3 \,{\rm TeV},\quad\
 \Lambda_{NP}\gsim\sqrt{|z^K_5|} \,10\times 10^3 \,{\rm TeV},
\end{split}
\ee
and
\be
\begin{split}\label{eq:4ferboundsD}
 \Lambda_{NP}&\gsim\sqrt{|z^D_1|}\, 1.2\times 10^3 \,{\rm TeV},\quad
 \Lambda_{NP}\gsim\sqrt{|z^D_2|} \,3.1\times 10^3 \,{\rm TeV},\quad
 \Lambda_{NP}\gsim\sqrt{|z^D_3|} \,1.6\times 10^3 \,{\rm TeV},\\
 \Lambda_{NP}&\gsim\sqrt{|z^D_4|} \,6.2 \times 10^3 \,{\rm TeV},\quad
 \Lambda_{NP}\gsim\sqrt{|z^D_5|} \,3.5\times 10^3 \,{\rm TeV}.
\end{split}
\ee
The above bounds are obtained after running $\alpha_3$ up to
$\Lambda_{NP}$ assuming that only SM fields contribute to the
beta-function up to that scale. In our models we will always have
additional fields below the scale $\Lambda_{NP}$, which will change
the running contribution to these bounds, but since the bounds are
up to unknown order one coefficients in any case, we ignore this issue here.
For a strongly coupled theory with
generic flavor structure one expects $|z_I^{K,D}|, |\tilde
z_I^{K,D}|\sim 1$ (barring accidental cancelations), and the above
inequalities directly translate into lower bounds on $\Lambda_{NP}$.

In supersymmetric extensions of the SM with perturbative
quark-squark-gluino interactions, the operators \eqref{eq:4ferops}
will also be generated through box diagrams with squark-gluino
exchange \cite{Hagelin:1992tc,Gabbiani:1996hi}, and in this case the
coefficients $z_I^{K,D}, \tilde z_I^{K,D}$ will be suppressed by
SM gauge couplings. Neglecting $\tilde q_L-\tilde q_R$ mixing, they can
be written as
\be\begin{split}\label{zis}
z_1^{K,D}&=-\frac{\alpha^2_3}{216}\left(24xf_6(x)+66\tilde
f_6(x)\right)(\delta_{12}^{d,u})^2_{LL},\\
\tilde z_1^{K,D}&=-\frac{\alpha^2_3}{216}\left(24xf_6(x)+66\tilde
f_6(x)\right)(\delta_{12}^{d,u})^2_{RR},\\
z_4^{K,D}&=-\frac{\alpha^2_3}{216}\left(504xf_6(x)-72\tilde
f_6(x)\right)(\delta_{12}^{d,u})_{LL}(\delta_{12}^{d,u})_{RR},\\
z_5^{K,D}&=-\frac{\alpha^2_3}{216}\left(24xf_6(x)+120\tilde
f_6(x)\right)(\delta_{12}^{d,u})_{LL}(\delta_{12}^{d,u})_{RR},
\end{split}\ee
where $f_6(x)$ and $\tilde f_6(x)$ are kinematical functions whose
expressions can be found in \cite{Gabbiani:1996hi} and $x$ is the ratio
of the gluino mass squared to the squark mass squared, $x=\tilde
m_{\tilde g}^2/\tilde m_{\tilde q}^2$. Flavor violation is encoded
in the mass insertion \cite{Hall:1985dx} parameters
\be\label{delta} (\delta^q_{12})_{NN}\equiv (K^q_{21})_N
(K^{q*}_{22})_N\,(\tilde m_{q_{N_2}}^2-\tilde m_{q_{N_1}}^2)/{\tilde
m_{\tilde q}^2}\,,~~~~(q=u, d,\ \ N=L,R) \ee
In the above expression the average squark mass is taken to be
$\tilde m_{\tilde q}\equiv (\tilde m_{q_{1}} +\tilde m_{q_{2}})/2$
\cite{Raz:2002zx} and $K_N^q=V_N^q\tilde V_N^{q\dagger}$ ($N=L,R$),
where $V_N^q$ and $\tilde V_N^q$ are hermitian matrices that
diagonalize the quark and squark mass matrices (again, neglecting
$\tilde q_L-\tilde q_R$ mixing), respectively, as
\be\label{eq:qdef} {\rm diag}(m_{q_1},m_{q_2},m_{q_3})= V_L^q M_q
V_R^{q\dagger},\ ~~~~ {\rm diag}(\tilde m^2_{ q_{N_1}},\tilde
m^2_{q_{N_2}},\tilde m^2_{q_{N_3}})=\tilde V_N^q \tilde M_{q_{NN}}^2
\tilde V_N^{q\dagger}. \ee

With this nomenclature, non-degenerate masses correspond to $(\tilde m_{q_{N_2}}^2-\tilde
m_{q_{N_1}}^2)\sim \tilde m^2_{\tilde q}$ and the amount of
alignment can be estimated from the size of the mixing angles
$(K^q_{21})_N (K^{q*}_{22})_N$. Generic squark mass matrices for the
first two generations with no degeneracy and no alignment lead to
$(\delta^q_{12})_{NN}\sim1$. To assess the amount of flavor
violation introduced by SUSY breaking in our subsequent (two-scale)
models, we will use $\alpha_3({\rm 1\ TeV})\simeq 0.089$ in
\eqref{zis}, set $\Lambda_{NP}\sim \tilde m_{\tilde q}$ in
\eqref{QCDH} with $z_I^{K,D}, \tilde z_I^{K,D}$ given by \eqref{zis}
and require that the total sum of the contributions proportional to
a single $\delta$ does not exceed the experimental values of $\Delta
m_{K,D}$.

\section{Fully coupled models}\label{sec:models}

In this paper we wish to explore the possibility that the same
dynamics generating the weak-scale flavor hierarchy is also
responsible for SUSY breaking, and to work out the qualitative form
of the resulting spectrum. Our basic framework is an implementation
of the Nelson-Strassler scenario as reviewed in Section
\ref{ssec:revNS}, leading to a suppression of the Yukawa couplings
through coupling of the first two generations to a strongly coupled
CFT sector as in \eqref{yukyuk}, but with SUSY dynamically broken at
a low scale by the CFT sector itself. This implies that the first
two generation superfields feel strong direct SUSY breaking effects;
note that this is very different from the high-scale SUSY breaking
discussed in \cite{Nelson:2001mq,Kobayashi:2001kz}.\footnote{In
these works SUSY breaking occurs at a high scale well above the
bottom of the conformal window, and so the dynamics of the conformal
regime leads to the suppression of many soft terms. There must then
be a long enough RG flow so that gauginos can drive the soft masses
back up to acceptable values, and so the scale $M_<$ is compelled to
be high.} In our scenario the standard model gauge group $G_{SM}$
necessarily couples (weakly) also to the CFT sector, so
gauge-mediated soft terms naturally also arise.

The above couplings are summarized in the following interaction Lagrangian:
\be \label{Lint} {\mathcal
L}_{int}=\left(\int\,d^2\theta\,\left(\lambda_1{\mathcal
O_1}\Phi_{1} + \lambda_2{\mathcal O}_2\Phi_{2}
+W({\mathcal O}_1,{\mathcal O}_2)\right)+
{\rm h.c.}\right)+\sum_{A=1}^3 2g_A\int d^4\theta\,{\mathcal
J}^A{\mathcal V}^A,
\ee
where the chiral operators ${\mathcal
O}_{1,2}$ have different and definite R-charges, $\Phi_{1,2}$ denote
generic first two generation MSSM matter superfields, the
couplings $\lambda_{1,2} \sim 1$ and we disregard irrelevant couplings following the discussion in Section \ref{ssec:revNS}.
${\mathcal V}^A$ are MSSM vector superfields and
${\mathcal J}^A$ are the CFT sector global symmetry current
superfields for $G_{SM}$ \cite{Meade:2008wd}.
Note that, in contrast to models of general gauge mediation, taking
$\alpha_A\rightarrow 0$ in our extended scenario does not fully
decouple the MSSM from the CFT sector, since the direct couplings
$\lambda_i$ are dynamically set by the superconformal theory.

We define $M$ as the mass of the degrees of freedom of the CFT
sector which couple directly to the first two generations. We assume
$M$ to be close to the bottom of the conformal window $M \sim M_<$,
although this assumption is easily relaxed. The scale $M$ may or may
not be equal to the scale of the mass gap for all CFT sector fields.
We distinguish models according to the number of scales involved:
\begin{itemize}
\item {\it One-scale models}: models in which the SUSY breaking scale is also
given by $M$.
\item {\it Two-scale models}: models where the
SUSY breaking scale $F$ is parametrically smaller than the
scale of decoupling and conformal symmetry breaking, $F\ll M^2$.
\end{itemize}

In the class of models that we are discussing, the Nelson-Strassler
mechanism necessitates that the CFT sector and SUSY breaking are not
flavor blind. This is simply because the first two generations
couple differently to the CFT sector. The main qualitative
constraint on our models therefore comes from the tension between
requiring small FCNCs, which pushes the mass of the CFT sector and
the first two generation sfermions upwards, and requiring small
fine-tuning for electroweak symmetry breaking, pushing the mass
scales of the theory downwards.

Our main result will be a class of models which are technically
natural and -- despite direct couplings between the first two
generations and the strongly coupled sector -- accommodate the FCNC
constraints without too much fine-tuning. The best models do so as a
result of being separable rather than non-separable, two-scale
models rather than one-scale models, and partially coupled rather
than fully coupled.

In this section we consider fully coupled models and their
limitations.  In Section \ref{ssec:M} we discuss the dynamics
associated with the scale $M$, in particular the constraints coming
from four-Fermi terms generated at the threshold scale $M$. Through
constraints on FCNCs we obtain a lower bound on $M$ assuming the
theory is generic (non-separable). We then show how a technically
natural structural requirement on the theory, which we call
separability, reduces the bound on $M$. Section \ref{sssec:soft}
discusses the soft terms. In Section \ref{ssec:1s} we address
one-scale models, where the same scale $M$ sets the soft masses in
the theory, leading to large fine-tuning.  In Section \ref{ssec:2s}
we discuss two-scale models, in which the mass of the first two
generation sfermions is reduced compared to one-scale models.
Although reducing the fine-tuning, this also introduces new sources
of FCNCs which arise due to the running of the first two generation
sfermions in loops \cite{Hagelin:1992tc,Gabbiani:1996hi}. Therefore
the fine-tuning issue is not fully resolved, but it is brought to a
much lower level than the single scale case.

\subsection{Flavor restrictions on $M$}
\label{ssec:M}
%

\subsubsection{Non-separable vs. separable models}\label{sssec:sep}

We are interested in low-scale SUSY breaking, leading to a spectrum
compatible with a light enough Higgs without fine-tuning. However,
generically, exiting the conformal and SUSY-preserving phase through
some strong coupling dynamics can be accompanied by troublesome
flavor-violating wave-function renormalizations (the graceful exit
problem) and four-Fermi operators. If generic four-Fermi terms of
the first two generations are generated at $M\sim M_<$, then the
bound on $M$ is the largest of the bounds in
Section~\ref{ssec:revFCNC},  $M\gsim~17~\cdot~10^3$~TeV. This leads
to a very heavy sparticle spectrum, resulting in large fine-tuning
in the Higgs sector. A concrete mechanism suppressing such effects
must be provided.

A possible resolution is the suppression of direct flavor-mixing
terms involving the first two generations in the CFT basis. One way
to achieve this is by using horizontal symmetries under which the
CFT sector fields are charged. Another way to accomplish this is by
assuming a ``decomposable'' CFT sector.  In such a setup, the CFT
sector decomposes into two separate sectors, each of which couples
to a specific linear combination of the three MSSM generations. In
particular, the gauge group $\tilde G$ decomposes into the product
of two groups, and the fields charged under one couple to those
charged under the other only via SM interactions.  In both
constructions terms similar to ${\mathcal O}_2\Phi_1$ can still
exist but, as discussed in Section \ref{ssec:revNS}, they are
negligible and do not affect the Yukawa structure. We will
henceforth omit these flavor mixing, R-symmetry violating terms from
the discussion; their effect on the flavor structure of the soft
terms will be at most comparable to the effects we discuss. We shall
refer to CFT sectors which are either decomposable or have
horizontal symmetries as ``separable'', and restrict ourselves to
such scenarios from now on.

Our separable CFT sectors give diagonal wave-function
renormalizations in the interaction basis, which realize the
graceful exit in a natural way. Nevertheless, even in the case of
separable models there can still be dangerous FCNC effects. These
come about from four-Fermi operators which appear flavor-diagonal in
the interaction basis, but are identified as flavor-changing when
switching to the mass basis.  (These effects are generically much
larger than the couplings induced by the SM interactions between the
two sectors.) Constraints from such operators, suppressed by powers
of $M$ as well as mixing angles, are still in tension with low
SUSY-breaking masses, as discussed in the following subsections.

\subsubsection{Separable CFT sectors}\label{ssec:sephid}

We now discuss the constraints on separable CFT sectors. In the
following we focus on decomposable CFT sectors; the case of
horizontal symmetries gives similar results.

We couple the first two generations of the MSSM to a decomposable
superconformal sector $\mathcal H$ with group $G\times \tilde
G_{1}\times \tilde G_{2}$. The gauge subgroups $\tilde G_1$ and
$\tilde G_2$ become strongly coupled at scales $\Lambda_1$ and
$\Lambda_2$, respectively. The interaction Lagrangian \eqref{Lint}
becomes
\be \label{THSL} {\mathcal
L}_{int}=\left(\int\,d^2\theta\,\left(\lambda_1{\mathcal
O}_1\Phi_1+\lambda_2 {\mathcal O}_2\Phi_2 + W_1({\mathcal O}_1) + W_2({\mathcal O}_2)\right)+{\rm
h.c.}\right)+\sum_{A=1}^32g _A\int\,d^4\theta\,{\mathcal
J}^A\,{\mathcal V}^A, \ee
where now the fundamental fields forming the composite operators
$\mathcal O_{1}$, $\mathcal O_2$ are not charged under $\tilde
G_{2}$, $\tilde G_1$, respectively, and there are no direct
couplings between fields charged under $\tilde G_1$ and $\tilde
G_2$. We collectively denote the CFT sector fields charged under
$\tilde G_{1,2}$ by ${\mathcal H}_{1,2}$. An example of this type of
models (although not SUSY breaking itself) is the example reviewed
in Section \ref{SU3} \cite{Nelson:2000sn}.

In these models each part of the CFT sector is perturbed away from
the fixed point by separate relevant deformations, and no new
intergenerational couplings are introduced.  We then naturally
obtain a flavor diagonal exit from the conformal regime. SUSY
breaking can occur in either sector separately, or in both. We will
assume SUSY is broken in both sectors, and for simplicity we take
the mass scales of ${\mathcal H}_{1,2}$ to be of similar order of
magnitude $M$.

The leading non-renormalizable interactions of the SM fields are
given by dimension six four-Fermi operators as in \eqref{NP}.
Schematically,
\be \label{genfourfermi}
 \mathcal{L}_{4-fermi} = \frac{1}{\Lambda_{NP}^2}\left(\sum_{I=1}^5 z_I
 {\mathcal Q}_I^{q_k q_l} +\sum_{I=1}^3 \tilde z_I\tilde {\mathcal Q}_I^{q_k q_l}\right),
\ee
and the constants $z_I$ and $\tilde z_I$ depend on the details of
the strong dynamics. Operators involving the first two generations
are suppressed by $\Lambda_{NP}\sim M$. In the absence of any
accidental cancelations in the coefficients, the first two
generations' terms have $z_I,\tilde z_I~\sim~{\cal O}(1)$, while terms
involving the third generation are further suppressed by powers of
SM gauge couplings. Since in separable models the two parts of the
CFT sector talk to each other only via SM gauge interactions, in the
CFT basis all first two generation four-Fermi operators are
flavor-diagonal at leading order. Dangerous flavor violating terms
are generated, however, when rotating to the mass basis
\eqref{eq:qdef}:
\beq\label{eq:4fermbasis}
\bar{\hat q}_{M_{_i}} \hat q_{N_{_i}} \bar{\hat q}_{P_{_i}} \hat q_{S_{_i}} \supset\left(\bar q_{M_{_1}} q_{N_{_2}} \bar q_{P_{_1}} q_{S_{_2}}\right)\left( V^{M}_{1i}V^{N\dagger}_{i2}V^{P}_{1i}V^{S\dagger}_{i2}\right)
 +\left(\bar q_{M_{_2}} q_{N_{_1}} \bar q_{P_{_2}} q_{S_{_1}}\right)\left( V^{M}_{2i}V^{N\dagger}_{i1} V^{P}_{2i}V^{S\dagger}_{i1}\right),
\eeq
where in the above $q=u,d$; $i=1,2$; $M,N,P,S = R,L$; the hatted
states on the left-hand side denote quarks in the CFT basis, while the
unhatted states on the right-hand side denote quarks in the mass
basis.

A bound on $M$ can now be obtained by combining
\eqref{eq:4fermbasis} with \eqref{eq:NSdiagmat}, \eqref{eq:4ferops},
\eqref{eq:4ferbounds} and \eqref{eq:4ferboundsD}. We find that the
strongest constraint comes from ${\cal Q}_4^{sd}$ in the
$K^0-\overline{K^0}$ system, which picks up a factor of $\sqrt{
m_d/m_s}\sim0.22$ compared to the general CFT sector estimate
$M\gsim 17\cdot 10^3$~TeV, namely (up to order one numbers coming
from the coefficients $z_I$ and ${\tilde z}_I$ in
\eqref{genfourfermi})
\beq\label{eq:Mbound2hid}
M \gsim 3.7\cdot 10^3\ \mathrm{TeV}.
\eeq

In models with partial couplings to the CFT sector the FCNC constraints can be relaxed. Such models will be discussed in Section \ref{sec:1c}.

\subsection{The soft terms in our models}\label{sssec:soft}
In our models there are two mechanisms for transmitting SUSY breaking
to the standard model fields. The first two generation superfields couple directly to
the CFT sector and obtain soft terms directly through these strong couplings. All other
fields of the standard model couple to the CFT sector only indirectly, mostly
through their interactions with the standard model gauge fields.  These couple directly,
though weakly, to the CFT sector, since the SM gauge group $G_{SM}$ is a
subgroup of the CFT sector global symmetry group. There are also very small contributions from the
Yukawa couplings of these fields to the first two generations, which we will ignore here.
The mechanism for generating these soft terms is thus gauge mediation through coupling to a strongly coupled theory.

In a theory of gauge mediation, the leading order (in standard model
gauge couplings) soft terms may be expressed in terms of correlators
of the $G_{SM}$ currents in the CFT sector theory
\cite{Meade:2008wd}. Since our CFT sector is strongly coupled, we
assume that these correlation functions are all of order one, up to
an overall factor of $N_{eff}^{(A)}$ which captures the effective
number of degrees of freedom in the CFT sector that are charged
under the $A$'th factor in the standard model gauge group and
participate in SUSY breaking. In a weakly coupled CFT sector this
would simply be the number of messenger fields, while in strongly
coupled theories it does not have to take integer values. We denote
by $\mu_S$ the scale at which the SUSY breaking dynamics is
integrated out, and by $\Lambda_S$ the effective SUSY breaking scale
transmitted to the MSSM (in one-scale models this is just the scale
$M$, but we will later analyze models where this scale is
different). Our assumptions then imply that at the scale $\mu_S$ the
gaugino masses, sfermion masses-squared (for multiplets which do not
couple directly to the CFT sector) and $A$-terms are
\be\label{eq:GGMmasses}
 M_A=
N_{eff}^{(A)}\frac{\alpha_A}{4\pi}\Lambda_S,\qquad \tilde
m_{f}^2\sim \sum_{A=1}^3
N_{eff}^{(A)}\left(\frac{\alpha_A}{4\pi}\right)^2 \Lambda_S^2,\qquad
A^{u,d,\ell}_{ij}\sim y^{u,d,\ell}_{ij} \sum_{A=1}^3
N_{eff}^{(A)}\left(\frac{\alpha_A}{4\pi}\right)^2 \Lambda_S, \ee
where $i,j$ are generation indices, $y^{u,d,\ell}$ are Yukawa
couplings, and the sum over $A$ should include only the SM gauge
groups that the specific fields in question are charged under. Note
that we define $N_{eff}^{(A)}$ in terms of the corresponding gaugino
mass.

In the equations above we ignored possible contributions from a D-term of the
hypercharge current coming from the CFT sector \cite{Fischler:1981zk,Dimopoulos:1996ig}. In gauge mediation
this D-term comes from a vacuum expectation value for the lowest component of the
$U(1)_Y$ current superfield, $J_Y$ \cite{Meade:2008wd},
and at tree-level it gives additional possible contributions to the
sfermion masses going as $\tilde m_{f}^2 = g_1^2 Y_f \langle J_Y \rangle$, where
$Y_f$ is the hypercharge of the sfermions. Naively we expect $\langle J_Y \rangle \sim
\pm N_{eff}^{(1)} \Lambda_S^2 / (4\pi)^2$, yielding contributions to sfermion masses of order
$\tilde m_f^2 \sim \pm Y_f N_{eff}^{(1)}\frac{\alpha_1}{4\pi}\Lambda_S^2$.  Such terms are larger than the
other contributions to third generation sfermion masses, and since they always take both
positive and negative values, they are problematic.

Thus, we will always assume that $U(1)_Y$ is embedded in a
non-Abelian group in the global symmetry $G$ of the CFT sector. This
means that at leading order we actually have $\langle J_Y \rangle =
0$, and the dominant contribution to $\langle J_Y \rangle$ comes
from the leading effects breaking this non-Abelian symmetry, which
we expect to be one-loop effects of order $\alpha_3 / 4\pi$
\cite{Dimopoulos:1996ig} (for instance, if it is a GUT symmetry).
The D-term contributions to the sfermion masses are
 then expected to be of order $\tilde m_f^2 \sim \pm Y_f N_{eff}^{(1)}\frac{\alpha_1\alpha_3}{(4\pi)^2}\Lambda_S^2$, which
 is already of the same order as the other gauge-mediated contributions \eqref{eq:GGMmasses} described above. We will
 analyze the effect of these terms in each scenario separately. In two-scale models, additional contributions to the D-term
 from the first two generation sfermions can arise via one-loop RG evolution, the effect of which will be discussed in Section \ref{ssec:2s}.

Having outlined some generalities of the soft terms structure, we
move on to discuss in detail separable one-scale and two-scale
models.

\subsection{Separable one-scale models}\label{ssec:1s}

In these models the scale of SUSY breaking is the same as that of
conformal symmetry breaking, and therefore all dimensionful
quantities are given in terms of powers of $M$. Specifically, the
soft terms for the first two generations are dominantly given by
\be\label{eq:bc121h1s}
(\tilde M_{u,d,\ell}^2)_{NN_{ii}} \sim M^2, \quad A^{u,d,\ell}_{ii}\sim y^{u,d,\ell}_{ii}M, \quad i=1,2, \quad  N=R,L,
\ee
and off-diagonal terms are not generated at leading order due to
separability of the CFT sector.  Since their mass is of order $M$,
the first two generation sfermions can be viewed as part of the CFT
sector and are integrated out when writing the effective action
below $M$. The effective Lagrangian below the scale $M$ thus takes
the form:
\be
\mathcal L_{eff} = \sum_{i=1,2} i q_{i}
D_{\mu}\gamma^{\mu} \bar{q}_{i} + \int  d^4 \,\theta \Phi_3
e^{\mathcal V}\Phi^{\dagger}_3 + \int d^2\,\theta
W_{MSSM}^{\tilde q_{1,2}=0}  + \mathcal L_{soft}^{\tilde
q_{1,2}=0} + \mathcal L_{\geq 4},
\ee
where $q_i$ are the first two generation fermions, $\tilde{q}_i$ are
their superpartners, and the MSSM superpotential $W^{\tilde
q_{1,2}=0}_{MSSM}$ and the soft Lagrangian ${\mathcal
L}_{soft}^{\tilde q_{1,2}=0}$ do not include the first and second
generation sfermions.  As we are discussing separable models,
wave-function renormalizations below $M$ are diagonal. The soft
Lagrangian is given by
\be
\begin{split}
-\mathcal L_{soft}^{\tilde
q_{1,2}=0} =  \frac{1}{2} \left (M_A \lambda^A \lambda^A + c.c.\right ) & +  \left ( (\tilde M_q^2)_{NN_{33}} \tilde q_{N_3}^* \tilde q_{N_3} + A^{q}_{33} H_{u,d} \tilde q_{L_3} \tilde q_{R_3}  + c.c. \right )\\
& +\left[ m_{H_u}^2H_u^*H_u+m_{H_d}^2H_d^*H_d+\left(B_{\mu}H_uH_d+c.c.\right)\right]
\end{split}
\ee
where $N =R,L$ and $q =u,d,\ell$. The gaugino masses $M_A$,
third generation soft masses $(\tilde M^2_q)_{33}$, soft trilinear
terms $A^q_{33}$ and soft Higgs terms are all dominantly generated
through gauge mediation \eqref{eq:GGMmasses} with an effective SUSY
breaking scale $\Lambda_S\sim M$.

Substituting the lowest value of $M$ compatible with the four-Fermi
operator bound \eqref{eq:Mbound2hid} into the general formulas
\eqref{eq:GGMmasses} and \eqref{eq:bc121h1s}, one finds that the
spectrum of such one-scale models is very heavy.  For instance, the
gluino mass is of order $22N^{(3)}_{eff}\ {\rm TeV}$ at the scale
$M$, and grows as it evolves down to the weak scale.  Clearly, a
spectrum of such massive gauginos and third generation sparticles is
unacceptable since it implies severe fine-tuning of the weak scale
\cite{Barbieri:1987fn,Dimopoulos:1995mi,Wright:1998mk}.  For this
reason we discard further discussion of these one-scale models, and
turn to models involving two scales.

\subsection{Separable two-scale models}\label{ssec:2s}

We now study models in which the SUSY breaking scale $F$ is
parameterically suppressed, $F \ll M^2$.  For simplicity, we discuss models in which
SUSY breaking occurs in both $\mathcal H_1$ and $\mathcal H_2$,
assuming $F_1\sim F_2\sim F$.

\subsubsection{Soft terms}\label{sssec:soft2s}

The soft terms for the first two generation sfermions can be
determined to leading order in $F/M^2$ by dimensional analysis and
by requiring these leading contributions to vanish in the
supersymmetric limit $F \rightarrow 0$, as well as in the limit of
no direct coupling $M \rightarrow \infty$. Chirality-preserving
diagonal soft masses are dominantly given by
\beq\label{eq:m12gen2s2h}
(\tilde M_{u,d,\ell}^2)_{NN_{ii}}\sim \left(\frac{F}{M}\right)^2,\quad i=1,2,\quad N=R,L.
\eeq
Off-diagonal soft masses-squared for the first two generations are
not generated from direct couplings, and are zero at leading order
in gauge mediation. The first two generation diagonal $A$-terms are given by
\begin{equation} \label{eq:A12}
A^{u,d,\ell}_{ii} \sim  y^{u,d,\ell}_{ii}\frac{F}{M},\ \ \ i=1,2,
\end{equation}
while off-diagonal first two generation $A$-terms are much smaller (see below).
In this scenario the first two generation sfermions are much lighter
than the CFT sector fields, so the parametric suppression of the
SUSY breaking order parameter $F/M^2 \ll 1$ can account for lighter masses
$\tilde m_{1,2}$ even for a large CFT sector scale $M$.

The third generation and gaugino soft masses are generated through gauge mediation, as are additional soft trilinear couplings.
The precise definition of $\Lambda_S$, the effective scale of SUSY breaking transmitted to the visible sector, in terms of $F$ and $M$
depends on the specific class of models under consideration, as we
will discuss shortly.  At this point, the only requirement is that
$\Lambda_S \rightarrow 0$ as $F \rightarrow 0$. For example, in
weakly coupled messenger models of gauge mediation $\Lambda_S$
corresponds to $F/M$, with $M$ the messenger scale and $F$ the
vacuum energy squared.  In terms of this scale $\Lambda_S$, we then have at
the scale $\mu_S$:
\beq\label{eq:GGM3}
\begin{split}
& M_A = N_{eff}^{(A)} \frac{\alpha_A}{4\pi} \Lambda_S,\\
& (\tilde M_{u,d,\ell}^2)_{NN_{33}}\sim N_{eff}^{(A)}
\left(\frac{\alpha_{A}}{4\pi}\right)^2 \Lambda_S^2
\pm Y_{u,d,\ell_{N}}N_{eff}^{(1)}\frac{\alpha_{1}\alpha_3}{(4\pi)^2} \Lambda_S^2,\quad N=R,L,\\
& A^{u,d,\ell}_{ij}\sim y^{u,d,\ell}_{ij}N_{eff}^{(A)} \left(\frac{\alpha_{A}}{4\pi}\right)^2 \Lambda_S,\quad i\neq j\quad{\rm or}\quad i=j=3, \quad i,j=1,2,3,\\
\end{split}
\eeq
where off-diagonal gauge-mediated soft masses-squared are
negligible, $Y_{u,d,\ell_{L,R}}$ denotes the hypercharge of the
appropriate sfermion, and in the above we take the largest
contribution when several SM gauge factors are possible.

In these models we need to consider the RG contribution of the first
two generation sparticles to the D-term \cite{Cohen:1996vb}.  The
soft masses-squared in \eqref{eq:m12gen2s2h} are more accurately
given by
\beq\label{eq:m12gen2s2hfull}
(\tilde M_{u,d,\ell}^2)_{NN_{ii}}\sim \left(\frac{F}{M}\right)^2 \pm \frac{\alpha_3}{4
\pi}\left(\frac{F}{M}\right)^2,\quad i=1,2,\quad N=R,L.
\eeq
The coefficient of the first term is assumed to be invariant under
$G$. The second term encodes the leading violation of the $G$
symmetry by the standard model. In the simplest case that $G$ is a
unified version of $G_{SM}$, the strongest such effect is
proportional to $\alpha_3$. Generally, the masses
\eqref{eq:m12gen2s2hfull} contribute to the D-term of $U(1)_Y$ and
to the soft masses of lighter particles. For some other light
sfermion of hypercharge $Y$, the D-term contributes
\be  Y \frac{\alpha_1}{4\pi}{\rm Tr}(Y_{f} \tilde m^2_{f}) \ee
to its soft mass-squared beta function. Depending on the sign of
$Y$, some of the third generation scalar masses may then be driven
tachyonic. Recall, however, that we assume that $U(1)_Y$ is embedded
in a non-Abelian factor in the global symmetry group $G$ of the CFT
sector which includes the first two generations (for example an
$SU(5)$ GUT). As a result, the first contribution from the heavy
first two generation masses in \eqref{eq:m12gen2s2hfull} yields a
vanishing hypercharge trace. Since the (say) GUT symmetry is broken
by SM gauge couplings, the second term in \eqref{eq:m12gen2s2hfull}
will induce a non-zero D-term, leading to contributions to light
sfermion mass-squared beta functions of order
\be \label{eq:D-term} Y \frac{\alpha_1}{4\pi}{\rm Tr}Y_{f} \tilde m^2_{f}\sim
Y\frac{\alpha_1\alpha_3}{(4\pi)^2}\left(\frac{F}{M}\right)^2. \ee
In the scenarios considered in the rest of the paper the integral of
\eqref{eq:D-term} is always smaller than the first contribution to
the sfermion masses-squared in \eqref{eq:GGM3}, and will henceforth
be neglected. The second term in the sfermion masses-squared
\eqref{eq:GGM3} can be negative and therefore potentially dangerous,
so it may require mild tuning.  We will discuss the effect of this
D-term in further detail for partially coupled models in Section
\ref{sec:1c}.

Operators of dimension $d\geq 4$ can also be generated.  Since
dimension four sfermion operators can only come from SUSY breaking,
they are parametrically suppressed via $F/M^2\ll1$, and are thus not
a concern.  In dimension five fermion-sfermion operators the SUSY
preserving effects dominate over the SUSY breaking ones.  The
four-Fermi operators of dimension six are dominated by the SUSY
preserving effects, and are down by $1/M^2$.  They dictate the bound
\eqref{eq:Mbound2hid} on $M$.

Equation \eqref{eq:m12gen2s2h} presents the pleasant feature of two-scale
models: the first two generation soft terms and the higher-dimensional operator coefficients can simultaneously
be suppressed by taking an enhanced value of
$M$ for fixed $F$, relaxing the contributions to FCNCs while lowering the scale of the scalar masses.

We consider the following two classes of models:

\subsubsection*{Class (a)}\label{sssec:1h2sa}

In these models the CFT sector is assumed to have a mass gap of order $M$, and all
of the CFT sector physics is integrated out at the scale
$M\sim\mu_S$. In the effective theory obtained at scale $M$
supersymmetry is dynamically broken, with an accidentally small SUSY
breaking parameter $F\ll M^2$. As a result
$\Lambda_S\sim F/M$.  In the limit $M \rightarrow \infty$ the CFT sector decouples and SUSY is not broken.

\subsubsection*{Class (b)}\label{sssec:1h2sb}

In these models the physics of the CFT sector coupling to the first
two generations is integrated out at the scale $M$ (namely, the CFT
sector fields that couple directly to the first two generations have
a mass of order $M$). The rest of the CFT sector is integrated out
at a lower scale $\sqrt F\ll M$, at which SUSY is then naturally
broken in the effective theory. Here the mass gap in the CFT sector
is the same as the SUSY breaking F-term. As a result
$\mu_S\sim\Lambda_S\sim\sqrt F$, while the soft terms for the first
two generations are still given by \eqref{eq:m12gen2s2h}.  In this
class of models the limit $M \rightarrow \infty$ corresponds to a
gauge mediation scenario with SUSY breaking  scale $\sqrt{F}$.
%
%
\vskip 20pt Since in models of Class (b) part of the CFT sector
dynamics is integrated out at $M$, we expect the SUSY breaking
dynamics in this class to yield smaller values of $N_{eff}$ than in
Class (a). Given the direct couplings and the MSSM quantum  numbers
of the operators ${\mathcal O}_{1,2}$ we expect $N_{eff}$ to be at
least twice the MSSM contribution (in two generations) for Class (a)
models. In Class (b) models we can have lower $N_{eff}$, though
requiring dynamical SUSY breaking in the CFT sector typically means
that it cannot be smaller than $\sim$ 5.

\subsubsection{Further FCNC bounds on scales}\label{sssec:1h2sFCNC}

As described above, in both classes of two-scale models generic
four-Fermi operators  generated by the superconformal dynamics are
suppressed by $1/M^2$, and bound $M$ according to
\eqref{eq:Mbound2hid}.  The non-zero soft terms involving the first
two generations are set by $F/M$ and are of the form
\eqref{eq:m12gen2s2h}, and chirality mixing terms are negligible. At
leading order, sfermion mass-squared matrices for the first two
generations are then already diagonalized in the CFT basis at the
scale $\mu_S$.

\begin{figure}
\begin{center}
\includegraphics[scale=1.5]{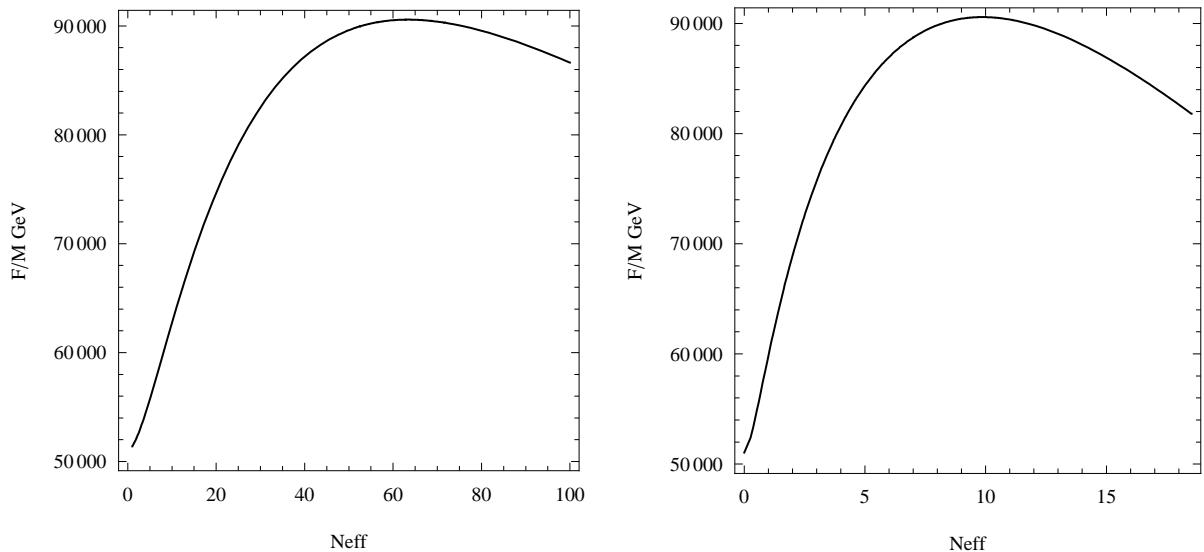}
\caption{\label{LLRRab} The relevant $K^0-\overline{K^0}$ system bounds
on $F/M$ as a function of $N_{eff}^{(3)}$ for fully coupled
two-scale models with a separable CFT sector ($M=3.7\cdot10^3 \,{\rm
TeV}$). The bounds depicted here come from
$(\delta^d_{12})_{LL}(\delta^d_{12})_{RR}$ in Class (a) (left) and
Class (b) (right) models.}
\end{center}
\end{figure}

In these models the sfermions are much lighter than $M$ and so
additional FCNC constraints are present. The most stringent bound on
$F/M$ is obtained from the $K^0-\overline{K^0}$ FCNC box diagrams
involving squarks and gluino exchange.  This  process constrains the
first two generation (down sector) masses at the weak scale,
obtained by RG evolution (RGE) from the boundary conditions
\eqref{eq:m12gen2s2h} at $\mu_S$ down to $m_Z$.  We can neglect RG
effects on the first two generation mass-squared matrices due to the
short range of scales involved, and use \eqref{eq:m12gen2s2h} at the
weak scale as well. In light of the above, in the squark sector we
have

\beq
(\tilde V^q_{L,R})_{ij}\sim \delta_{ij},\quad i,j,=1,2,\quad q=u,d.
\eeq
The quark sector diagonalizing matrices have the structure
\eqref{eq:NSdiagmat}, and so the mixing matrices in \eqref{delta}
are of the form
\be\label{eq:K}
(K^{q}_L)_{ij} \sim (V^q_L)_{ij}\sim |V_{ij}|,\ \ \ (K^{q}_R)_{ij} \sim (V^q_R)_{ij}\sim \frac{m_{q_i}/m_{q_j}}{|V_{ij}|}\quad (i<j),\quad i,j=1,2,\quad q=u,d.
\ee
Note that $(K^d_L)_{12}\sim |V_{12}|\sim(K^d_R)_{12}\sim
\frac{m_d/m_s}{|V_{12}|}\sim 0.22$ are all of order a Cabibbo
factor. In terms of the $\delta_{12}^d$ parameters of Section
\ref{ssec:revFCNC}, the separable CFT sector scenario with no
degeneracy then corresponds to taking all $\delta^d_{12}$ of order a
Cabibbo factor $\sim 0.22$.  Additionally, \eqref{eq:m12gen2s2h} and
\eqref{eq:GGM3} imply that the ratio $x\equiv m_{\tilde g}^2/\tilde
m_{\tilde q}^2$ appearing in \eqref{zis} is small yet depends on
$N^{(3)}_{eff}$.  In models of Class (a), this ratio is independent
of $F/M^2$ and is given by
\beq\label{eq:xa} x_{(a)}=\left(\frac{\alpha_3}{4\pi}
N_{eff}^{(3)}\right)^2 \eeq
while in models of Class (b) we have
\beq\label{eq:xb}
x_{(b)}=\left(\frac{\alpha_3}{4\pi}N_{eff}^{(3)}\right)^2\frac{M^2}{F}.
\eeq
The most stringent bounds come from the mixed
$(\delta^d_{12})_{LL}(\delta^d_{12})_{RR}$ terms in \eqref{zis}. We
present plots of these bounds as a function of $N_{eff}^{(3)}$ for
Class (a) and Class (b) models in Figure \ref{LLRRab}.
Representative bounds are\footnote{Recall that all the bounds we
write are up to unknown order one constants coming from correlators
in the CFT sector.}
\beq\label{eq:FoverM2c}
\begin{split}
&{\rm Class\ (a):}\ \ \ \frac{F}{M}\gsim\ 81\ {\rm TeV}\ \ \ (N^{(3)}_{eff}=27),\\
&{\rm Class\ (b):}\ \ \ \frac{F}{M}\gsim\ 87\ {\rm TeV}\ \ \
(N^{(3)}_{eff}=5).
\end{split}
\eeq
The values for $N_{eff}^{(3)}$ are chosen in agreement with the
two-loop effect \cite{ArkaniHamed:1997ab,Agashe:1998zz} that will be
discussed in Section \ref{sec:1c}. Flavor changing chirality-mixing
terms $(\delta^d_{12})_{LR}$ give much less stringent constraints on
$F/M$ due to the relative smallness of the $A$-terms \eqref{eq:A12}.

Given some amount of degeneracy in the first two generation down sector, the bound is weakened and lower values of $F/M$ are accessible.
 For example, allowing additional $\sim 0.5$ degeneracy
the bounds in both classes are weakened
 to $F/M\gsim 41$~TeV and $F/M\gsim 43$~TeV respectively, for the same values of $N^{(3)}_{eff}$ as in \eqref{eq:FoverM2c}.

Using the bounds on $M$ and $F/M$ given in \eqref{eq:Mbound2hid} and
\eqref{eq:FoverM2c}, we compute a bound for $\sqrt{F}$ relevant for
models of Class (b), yielding
\be\label{eq:Fbound} \sqrt{F} \gsim 570\ {\rm TeV}. \ee
It is clear from \eqref{eq:m12gen2s2h}, \eqref{eq:GGM3},
\eqref{eq:FoverM2c} and \eqref{eq:Fbound} that in the two-scale
models described above the generic sfermion soft mass spectrum at
the scale $\mu_S$ is still given by an inverted hierarchy, provided
$N_{eff}$ is not too large. In such models, avoiding third
generation tachyons from heavy first two generation two-loop
contributions \cite{ArkaniHamed:1997ab, Agashe:1998zz} puts a lower
bound on initial third generation masses, which in our models
translates into a lower bound on $N^{(A)}_{eff}$.  This two-loop
effect will be discussed in greater detail in Section \ref{sec:1c}.
In the current context, this effect along with the initial soft
terms \eqref{eq:GGM3} imply unnaturally large initial  soft masses
for the third generation sfermions and the gauginos. For instance, gluino masses
at the scale $\mu_S$ are of order $\sim 12$ TeV in Class (a), and
$\sim 18$ TeV in Class (b). These fully coupled separable models
thus still require large fine-tuning in the electroweak sector.


\section{Partially coupled models}\label{sec:1c}

In the models analyzed in the previous section all MSSM fields of
the first two generations coupled directly to the CFT sector.  This
resulted in relatively high scales $M$ and $F/M$, which led to large
fine-tuning of the weak scale $m_Z$. In this section we explore
models with partial couplings, in which the strongest constraints on
the scales can be relaxed and the fine-tuning can be alleviated.

When fully coupling all first two generation fields to the CFT
sector, the main sources of flavor violation are operators made out
of both left- and right-handed fields in the down sector.  We can
thus allow for a lower scale $M$ by coupling only some of these
fields (in particular only some of the down quarks) directly to the
CFT sector.  Here we will consider several such models -- models in
which only the left-handed (LH) fields in the first two generations
couple directly to the CFT sector, similar models with right-handed
(RH) direct couplings, and a {\bf 10}-centered model, in which only
the ${\bf 10}$'s of $SU(5)$ couple to the CFT sector, thus coupling
both up sector chiralities but suppressing mixed chirality-operators
in the down sector. We consider here only two-scale models, of both
Class (a) and (b).

A few comments concerning D-terms are in order. When coupling only a
single chirality to the CFT sector, unification is lost. In order
to suppress the tree-level gauge mediation and one-loop beta-function D-terms as in
the GUT case (see \eqref{eq:GGM3} and \eqref{eq:D-term}, respectively), one is forced to embed $U(1)_Y$ in some non-Abelian
group under which left and right chirality superfields transform separately \cite{Cohen:1996vb,Dimopoulos:1996ig}.
In this case one should require that the parameter $\zeta$ measuring
the magnitude of the breaking of the non-Abelian symmetry should be such that the one-loop gauge-mediated masses-squared
\be \label{eq:sfmch} \tilde m_{f}^2\sim \sum_{A=1}^3
N_{eff}^{(A)}\left(\frac{\alpha_A}{4\pi}\right)^2 \Lambda_S^2 \pm
Y_f N_{eff}^{(1)}\frac{\alpha_1\zeta}{(4\pi)^2}\Lambda_S^2 \ee
do not lead to tachyonic sfermions. As discussed in Section
\ref{sssec:soft}, the D-term contribution to the one-loop beta
function for the sfermion masses-squared \eqref{eq:D-term} (where
$\alpha_3$ is replaced by a general $\zeta$) is negligible in
comparison to the bare D-term and can be neglected from subsequent
discussion.\footnote{We implicitly assume here that $\zeta$ is not
larger (at the scale $M$) than GUT breaking parameters in the
standard model such as $\alpha_3$, and that $N_{eff}^{(1)}$ is not
much larger than the other $N_{eff}$'s.}

We will analyze the effect of the remaining gauge mediation D-term initial masses-squared in \eqref{eq:sfmch} for the
third generation sfermions case by case.

\subsection{The two-loop effect}\label{ssec:2loop}

As in fully coupled models, all the constructions discussed here
will exhibit a (partial) inverted hierarchy in the sfermion sector.
Inverted hierarchy models are known to be susceptible to two-loop
effects in the RGE, where the heavy first two generations can render
light third-generation sparticles tachyonic
\cite{ArkaniHamed:1997ab, Agashe:1998zz}.

Positive physical masses-squared then require heavy initial soft
masses for the third generation, such that a fair amount of
fine-tuning may be necessary to stabilize the weak scale. This
concern, however, will turn out to be unjustified -- some
fine-tuning is needed, but for a reasonable range of scales and
parameters it can be milder than, say, the percent level.

The beta function for the third generation (and all non-directly
coupled) sfermion masses-squared up to two-loops can be written as
\cite{Martin:1993zk}
\be\label{betafunction} \frac{d}{dt}\tilde{m}_{f}^2=\frac{1}{16
\pi^2}\beta^{(1)}_{\tilde m^2_{f}}+\frac{1}{(16
\pi^2)^2}\beta^{(2)}_{\tilde m^2_{f}},\ \ \ t\equiv
\log\left(\mu/\mu_0\right). \ee
In our analysis we will neglect Yukawa couplings and $A$-terms
throughout in the RGE, as well as gaugino contributions at the
two-loop level. With these approximations the beta-functions can be
written as \cite{Martin:1993zk}
\begin{eqnarray}\label{eq:stauR}
\begin{split}
\beta^{(1)}_{\tilde m^2_{f}} & \simeq -8 \sum_A g_A^2 C_A(R_{f}) |M_A|^2,  \\
\beta^{(2)}_{\tilde m^2_{f}} & \simeq  +4 \sum_A g_A^2 C_A(R_{f})  \sigma_A + \frac{12}{5}g_1^2
Y_{f} {\mathcal S}',
\end{split}
\end{eqnarray}
where the sum over $A$ runs over the three SM gauge groups,
$C_A(R_{f})$ denotes the quadratic Casimir invariant of the
representation $R_{f}$ of the gauge group $A$, $Y_{f}$ is the
hypercharge of the sfermion $f$, and (ignoring small contributions
from soft Higgs masses) one has
\begin{eqnarray}
\begin{split}
\sigma_1  \simeq & \frac{1}{5}g_1^2\mathrm{Tr}\left(\tilde{M}_{Q_L}^2+3\tilde{M}_{L_L}^2+8\tilde{M}_{u_R}^2+2\tilde{M}_{d_R}^2+6\tilde{M}_{e_R}^2\right),\\
\sigma_2  \simeq & g_2^2\mathrm{Tr}\left(3\tilde{M}_{Q_L}^2+\tilde{M}_{L_L}^2\right),\\
\sigma_3  \simeq &  g_3^2\mathrm{Tr}\left(2 \tilde{M}_{Q_L}^2+\tilde{M}_{u_R}^2 + \tilde{M}_{d_R}^2 \right),
\end{split}
\end{eqnarray}
and
\begin{eqnarray}\label{Sprime}
\begin{split}
\mathcal S' \simeq & \frac{8}{3} g_3^2 \mathrm{Tr}\left(\tilde{M}_{Q_L}^2-2\tilde{M}_{u_R}^2+\tilde{M}_{d_R}^2\right)+\frac{3}{2}g_2^2\mathrm{Tr}\left(\tilde{M}^2_{Q_L}-\tilde{M}^2_{L_L}\right)\\
&+g_1^2\mathrm{Tr}\left(-\frac{3}{10}\tilde{M}^2_{L_L}+\frac{1}{30}\tilde{M}^2_{Q_L}-\frac{16}{15}\tilde{M}^2_{u_R}+\frac{2}{15}\tilde{M}^2_{d_R}+\frac{6}{5}\tilde{M}^2_{e_R}  \right).
\end{split}
\end{eqnarray}
In the above equations we write the sfermion mass-squared matrices
in the standard $SU(2)$ notation.  In all the subsequent estimates
made in this section, \eqref{betafunction} is integrated in the
leading $\log$ approximation for the appropriate light sparticles,
where the two-loop contribution runs down to the approximate
decoupling scale for the heavy first two generation scalars $\sim
F/M$, and the one-loop term flows down to a scale $\mu_0$ $\sim 1$
TeV. Neglecting chirality-mixing, we then impose that for the
lightest sparticle
\begin{equation}\label{eq:RGEmass}
\tilde m_{f}^2(\mu_0)\simeq \tilde m_{f}^2(\mu_S) + \frac{1}{16 \pi^2}\beta^{(1)}_{\tilde m^2_{f}}(\mu_S)\log\left(\frac{\mu_S}{\mu_0}\right)+\frac{1}{(16 \pi^2)^2}\beta^{(2)}_{\tilde m^2_{f}}(\mu_S)\log\left(\frac{\mu_S}{F/M}\right)\gsim 0.
\end{equation}
Taking $N_{eff}^{(A)} \equiv N$ for simplicity in all initial
sfermion and gaugino masses, a rough lower bound on $N$ may then be
found in each scenario, affecting the initial soft terms and thus
the physical spectrum.  A full analysis will of course introduce
corrections to the obtained bounds, but the order of magnitude of
$N$ is captured in this approximation. The sample spectra that we
present in Section \ref{ssec:pheno} use the full two-loop RG
evolution, and are consistent with this estimate.

\subsection{Models}\label{ssec:partialmodels}

The ground is now set to explore some partially coupled models.  We begin with chiral models in which only the
MSSM fermion-sfermion fields of one chirality are coupled to the
superconformal sector.  We then address an example of non-chiral partial couplings -- a unified {\bf 10}-centered scenario.

\subsubsection{Chiral models}\label{sssec:1c}

We assume that only the left-handed (right-handed) MSSM fields are
coupled to the CFT sector.  Therefore, in the quark and lepton
sectors, there are only left-handed (right-handed) suppression
factors generated by the superconformal dynamics, and the
factorizable Yukawa structure of \eqref{yukyuk} contains a single
$\epsilon_{L}$ ($\epsilon_R$) factor.  These models can still lead
to acceptable flavor structures, but producing appropriate
suppression for given anomalous dimensions requires increasing the
ratio $M_>/M_<$.
There are, however, no {\it a priori} constraints on this
ratio.\footnote{We need to require that $M_>$ is below the Planck
scale, but this is easy to satisfy. Additional constraints from
Landau poles will be discussed in the next section.}  To be more
concrete, in chiral left-handed models, the suppression pattern
predicts that the mixing angles are of order the mass ratios, and so
some additional suppression is needed in order to be in full
agreement with measurement.  Right-handed models suffer from a
difficulty to parametrically produce the CKM mixing angles.  When
discussing right-handed models in the following, we assume a Cabibbo
factor has been generated in estimating the right-handed mixing
matrices \eqref{eq:NSdiagmat}; the bounds on the scales obtained in
this way are conservative.

A comment is in order. When only right-handed superfields are
coupled to the CFT sector, the minimal and more natural assumption
is that the CFT sector is uncharged under $SU(2)_L \subset G_{SM}$.
In such a scenario the Wino gauge-mediated mass will not be
generated at the high scale $\mu_S$, leading to unacceptably light
neutralinos and charginos at the weak scale. Therefore, from here on
when discussing chiral right-handed couplings we assume that the CFT
sector is charged under $SU(2)_L$. This also leaves open the
possibility of directly coupling the Higgs sector to the CFT sector.

Among the four-Fermi operators \eqref{genfourfermi} bounding $M$,
only the chiral operators ${\cal Q}_1$ (${\tilde{\cal Q}}_1$) are
now generated at $M$.  In light of \eqref{eq:NSdiagmat}, the bound
\eqref{eq:Mbound2hid} in separable models can then be relaxed here
to
\beq\label{eq:Mbound2h1c}
\begin{split}
&\mathrm{LH\ couplings:}\ \ \ \ M\gsim 264{\rm\ TeV}\ \ (\mathrm{from\ the\ }D^0-\overline{ D^0} \mathrm{\ system}),\\
&\mathrm{RH\ couplings:}\ \ \ \ M\gsim 220{\rm\ TeV}\ \
(\mathrm{from\ the\ }K^0-\overline{ K^0} \mathrm{\ system}).
\end{split}
\eeq

In the sfermion sector, soft mass-squared terms for the left-handed
(right-handed) first two generation sfermions are given by direct
coupling as in \eqref{eq:m12gen2s2h}
\beq\label{eq:m121c} (\tilde M^2_{u,d,\ell})_{LL(RR)_{ii}}\sim
\left(\frac{F}{M}\right)^2,\quad i=1,2\,. \eeq
All other soft terms are gauge-mediated similarly to \eqref{eq:GGM3}:
\beq\label{eq:GGM1c}
\begin{split}
&M_A= N_{eff}^{(A)} \frac{\alpha_A}{4\pi} \Lambda_S,\\
&(\tilde M_{u,d,\ell}^2)_{LL,RR_{33}}\sim N_{eff}^{(A)}
\left(\frac{\alpha_{A}}{4\pi}\right)^2 \Lambda_S^2
\pm N_{eff}^{(1)}Y_{u,d,\ell_{L,R}} \frac{\alpha_{1}\zeta}{(4\pi)^2} \Lambda_S^2,\\
&(\tilde M_{u,d,\ell}^2)_{RR(LL)_{ii}}\sim N_{eff}^{(A)}
\left(\frac{\alpha_{A}}{4\pi}\right)^2 \Lambda_S^2
\pm N_{eff}^{(1)}Y_{u,d,\ell_{R(L)}} \frac{\alpha_{1}\zeta}{(4\pi)^2} \Lambda_S^2\ ,\ \ i=1,2,\\
&A^{u,d,\ell}_{ij}\sim
y^{u,d,\ell}_{ij}N_{eff}^{(A)}\left(\frac{\alpha_{A}}{4\pi}\right)^2
\Lambda_S,\quad i,j=1,2,3.
 \end{split}
\eeq
Avoiding initial tachyonic right-handed sleptons when the sign of the
D-term is negative requires $\zeta \sim \alpha_3/5$, implying a mild
tuning of the D-term generated by the CFT sector in this case. In
Class (b) models we can achieve suppression of the
D-term naturally by assuming a messenger parity
\cite{Dimopoulos:1996ig} symmetry (taking $J_Y \rightarrow -J_Y$)
below the scale $M$. In such a case $\langle J_Y \rangle_{1-loop} =
0$ \cite{Meade:2008wd,Dimopoulos:1996ig} and so
\be
\langle J_Y \rangle = \mathcal O\left(\frac{\zeta^2}{16\pi^2}\right),
\ee
{\it i.e}, the D-term mass-squared acquires an extra factor of $\zeta/(4\pi)$.

In the chiral coupling scenarios discussed here $N_{eff}^{(A)}$
could be smaller than in the case of direct coupling of both
left- and right-handed fields to the superconformal sector,
schematically by a factor of a half.

In chiral left-handed models the most stringent bounds on the masses
of the heavy first two generations, $F/M$, are set by processes
involving the $(\delta_{12}^u)_{LL}$ term in $D^0-\overline{D^0}$ box
diagrams. The $(\delta_{12}^d)_{LL}$ term in $K^0-\overline{K^0}$ box
diagrams gives comparable yet milder constraints. Since the
right-handed soft masses-squared are given by gauge mediation there is near
degeneracy in the first two generations of right-handed squarks, and
so $(\delta^{u,d}_{12})_{RR}\approx0$ for left-handed models.

On the other hand, in chiral right-handed models, the strongest
constraints on the heavy scale $F/M$ are obtained from processes
involving $(\delta_{12}^d)_{RR}$ terms in $K^0-\overline{K^0}$ box
diagrams. Processes involving $(\delta_{12}^u)_{RR}$ terms in
$D^0-\overline{D^0}$ box diagrams are comparatively suppressed by
the relative smallness of the right mixing angles in the up sector
(see \eqref{eq:NSdiagmat}). Since the left-handed soft
masses-squared are given by gauge mediation there is near degeneracy
in the first two generations of left-handed squarks, and so
$(\delta^{u,d}_{12})_{LL}\approx0$ for right-handed models.

\begin{figure}
\begin{center}
\includegraphics[scale=1.5]{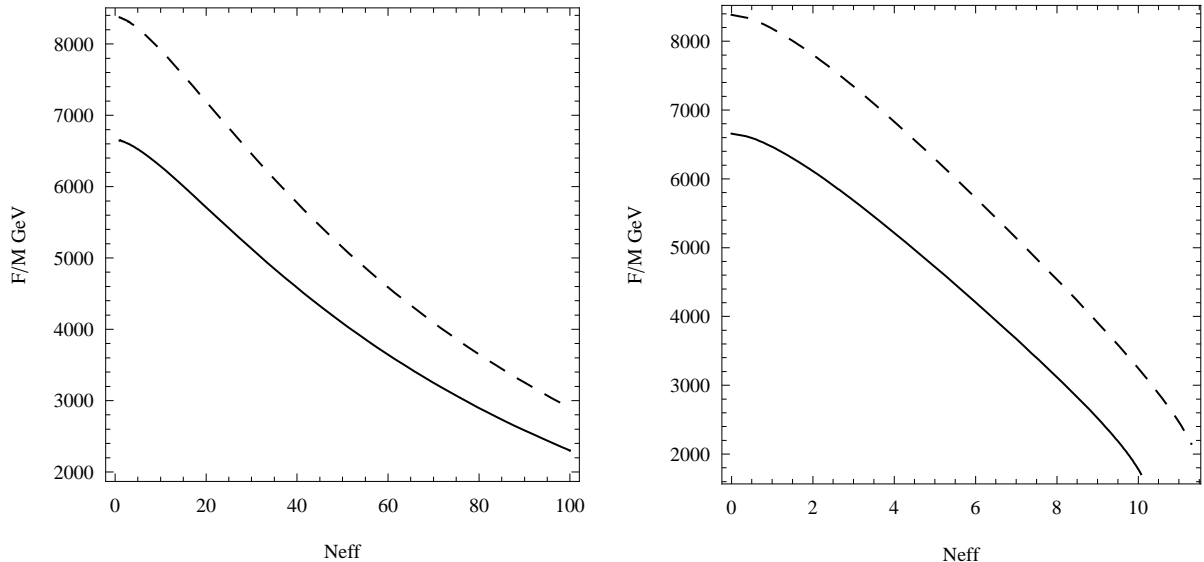}
\caption{\label{LHLLab} Relevant $D^0-\overline{D^0}$ system
(dashed) and $K^0-\overline{K^0}$ system (bold) bounds on $F/M$ as a
function of $N_{eff}^{(3)}$ for chiral left-handed models
($M=264\,{\rm TeV}$). The left graph refers to Class (a) models and
the right graph refers to Class (b) models.}
\end{center}
\end{figure}

\begin{figure}
\begin{center}
\includegraphics[scale=1.5]{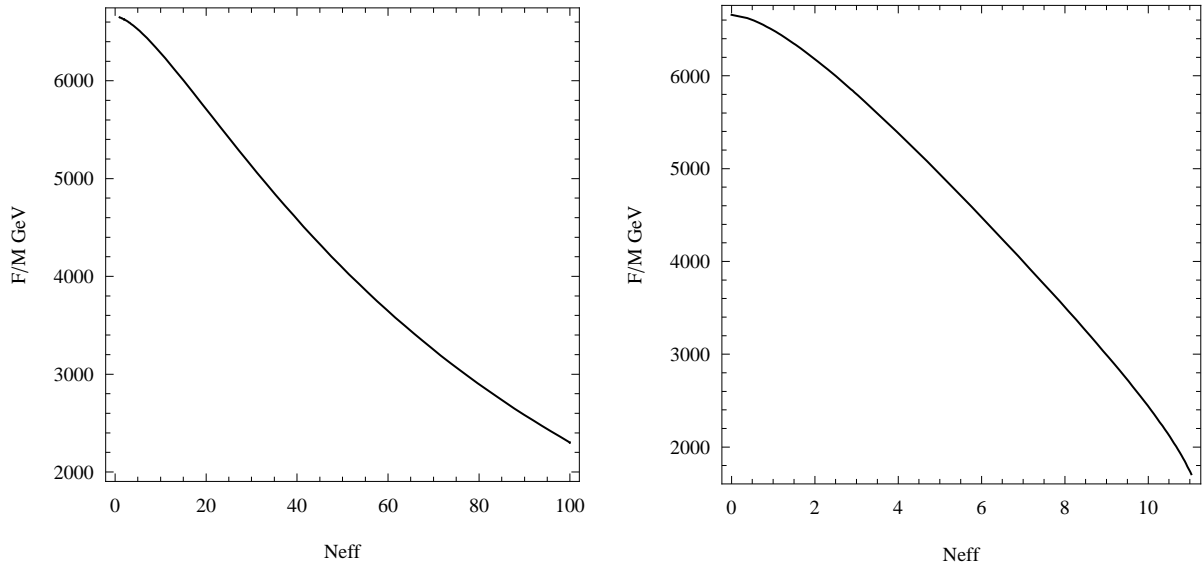}
\caption{\label{RHLLab} Relevant $K^0-\overline{K^0}$ system bounds on
$F/M$ as a function of $N_{eff}^{(3)}$ for chiral right-handed models
($M=220\,{\rm TeV}$). The left graph refers to Class (a) models and
the right graph refers to Class (b) models.}
\end{center}
\end{figure}

Plots of the bounds on $F/M$ as a function of $N_{eff}^{(3)}$ are given in
Figure \ref{LHLLab} for chiral left-handed models, and
in Figure \ref{RHLLab} for chiral right-handed models.

Representative bounds in chiral left-handed coupled models are
\beq\label{eq:FoverM1cLH}
\begin{split}
&{\rm LH\ Class\ (a):}\ \ \ \frac{F}{M}\gsim\ 5\ {\rm TeV}\ \ \ (N^{(3)}_{eff}=62),\\
&{\rm LH\ Class\ (b):}\ \ \ \frac{F}{M}\gsim\ 6\ {\rm TeV}\ \ \
(N^{(3)}_{eff}=5),
\end{split}
\eeq
while in chiral right-handed coupled models they are
\beq\label{eq:FoverM1cRH}
\begin{split}
&{\rm RH\ Class\ (a):}\ \ \ \frac{F}{M}\gsim\ 6\ {\rm TeV}\ \ \ (N^{(3)}_{eff}=15),\\
&{\rm RH\ Class\ (b):}\ \ \ \frac{F}{M}\gsim\ 5\ {\rm TeV}\ \ \
(N^{(3)}_{eff}=5).
\end{split}
\eeq
Combining the bounds on $M$ and $F/M$ of \eqref{eq:Mbound2h1c},
\eqref{eq:FoverM1cLH} and \eqref{eq:FoverM1cRH} we obtain for models
of Class (b):
\beq
\begin{split}
&\mathrm{LH\ couplings:}\ \ \ \sqrt F\gsim 40\ {\rm TeV},\\
&\mathrm{RH\ couplings:}\ \ \ \sqrt F\gsim 33\ {\rm TeV}.
\end{split}
\eeq

As explained in Section \ref{ssec:2loop}, two-loop contributions to
third generation masses put a lower bound on $N\equiv
N_{eff}^{(A)}$. For concreteness we obtain the bounds for the point
in parameter space where all order-one numbers in \eqref{eq:m121c}
are equal to one. The analysis for other points in parameter space
is qualitatively the same.  Using \eqref{eq:m121c} and
\eqref{eq:GGM1c} for left-handed models, at this point in parameter
space one has
\be
\begin{split}
& \mathcal S' \simeq  \left[\frac{16}{3}g_3^2-\frac{8}{15}g_1^2\right]\left( \frac{F}{M}\right)^2,\\
& \sigma_1  \simeq  \frac{8}{5}g_1^2\left( \frac{F}{M}\right)^2,~~\sigma_2  \simeq  8g_2^2\left( \frac{F}{M}\right)^2, ~~\sigma_3  \simeq  4g_3^2\left( \frac{F}{M}\right)^2.
\end{split}
\ee
For Class (a) the strongest bound on $N$ comes from the right-handed
charged sleptons, for which \eqref{eq:RGEmass} reads
\begin{eqnarray}\label{eq:2loop}
\begin{split}
\tilde{m}_{e_{Ri}}^2({\rm TeV}) \sim  \frac{\alpha_1 \Lambda_S^2}{(4 \pi)^2} &\left[\alpha_1 N+\frac{12}{5}\left(- \frac{16}{3}\alpha_3-\frac{16}{15}\alpha_1\right)\left(\frac{F/M}{\Lambda_S}\right)^2\log\left(\frac{\mu_S}{F/M}\right) \right.\\
 & \ \ \ \ \ \ \ \ \ \ \  +  \left. \frac{24}{5}\frac{\alpha_1^2}{4 \pi}N^2\log\left(\frac{\mu_S}{{\rm TeV}}\right)\right]\gsim 0,\ \ i=1,2,3.
\end{split}
\end{eqnarray}
For Class (b) the strongest bound comes from $\tilde u_{R_i}$,
although other sparticles give comparable bounds.\footnote{Here and in the following, since we neglect Yukawa
couplings in the running, all generations of the non-directly
coupled fields have similar RG evolution. Including the effects of Yukawas, the
strongest constraints will come specifically from third generation
sparticles within the relevant sector.} By solving the relevant
inequalities we obtain the bounds:

\be\label{eq:NLH}
\begin{split}
{\rm LH~Class\ (a)}:\ \ \ & N\gsim 62  \ \ \textrm{for} \  \
\Lambda_S \sim F/M=5\ {\rm TeV},\ \mu_S \sim M=264\ {\rm TeV},\\
{\rm LH~Class\ (b)}:\ \ \ & N\gsim 2 \ \ \textrm{for} \  \ \Lambda_S
\sim \mu_S\sim \sqrt{F}=40\ {\rm TeV}\ (F/M=6\ {\rm TeV}).
\end{split}
\ee
The analysis is analogous for right-handed models. For Class (a)
models all sparticles, except right-handed charged sleptons which do
not become tachyonic, give similar bounds, while in Class (b) models
all sparticles typically give comparable constraints.  We obtain:
\be\label{eq:NRH}
\begin{split}
{\rm RH~Class\ (a)}:\ \ \ & N\gsim 15  \ \ \textrm{for} \  \
\Lambda_S \sim F/M=6\ {\rm TeV},\ \mu_S \sim M=220\ {\rm TeV},\\
{\rm RH~Class\ (b)}:\ \ \ & N\gsim 2 \ \ \textrm{for} \  \ \Lambda_S
\sim \mu_S\sim \sqrt{F}=33\ {\rm TeV}\ (F/M=5\ {\rm TeV}).
\end{split}
\ee
The values of $N_{eff}^{(3)}$ in \eqref{eq:FoverM1cLH} and \eqref{eq:FoverM1cRH} are chosen in agreement with the above.

\subsubsection{{\bf 10}-centered models}\label{sssec:10}

As an alternative scenario, one could contemplate evading the
strongest $K^0-\overline{K^0}$ mixing bounds by disallowing
non-chiral couplings in the down sector alone. In $SU(5)$-based GUT
models, where the ${\bf \bar 5}$ contains $L_L, d_R$ and the ${\bf
10}$ contains $Q_L, u_R$ and $e_R$, coupling only the ${\bf
10}_{1,2}$ to the CFT sector can accomplish this, while still
generating the Yukawa hierarchies in the up, down and lepton sectors
and guaranteeing a vanishing hypercharge $D$-term at leading order.
In this scenario, suppression factors for both chiralities are
generated in the up quark sector. In the down quark and lepton
sectors only fields of one chirality, left and right respectively,
acquire large anomalous dimensions.  A detailed discussion of this
fermion flavor structure can be found in \cite{Nelson:2000sn}.

In this setup the strongest bound on $M$
comes from ${\cal Q}_2$ of \eqref{genfourfermi} in the $D^0-\overline{D^0}$ system \cite{Bona:2007vi} and reads
\beq M\gsim 682\ {\rm TeV}. \eeq
First two generation up squarks feel direct SUSY breaking, as do
left-handed down squarks and right-handed sleptons, while all other
sparticles have gauge mediated soft masses and masses-squared.
Diagonal first two generation soft trilinear couplings in the up
sector have direct SUSY breaking in them and similarly to
\eqref{eq:bc121h1s} are proportional to the appropriate Yukawa
couplings and are not suppressed by SM gauge factors.  All other
$A$-terms are dominantly given by gauge mediation.  The expressions
in \eqref{eq:m121c} and \eqref{eq:GGM1c} are appropriately modified.

As in chiral models, avoiding initial tachyonic right-handed
sleptons requires a tuning of order $1/5$ in the D-term for one of
the possible signs of its contribution to the soft masses-squared.
In Class (b) models, this tuning is not necessary if we assume
messenger parity below the scale $M$.

\begin{figure}
\begin{center}
\includegraphics[scale=1.5]{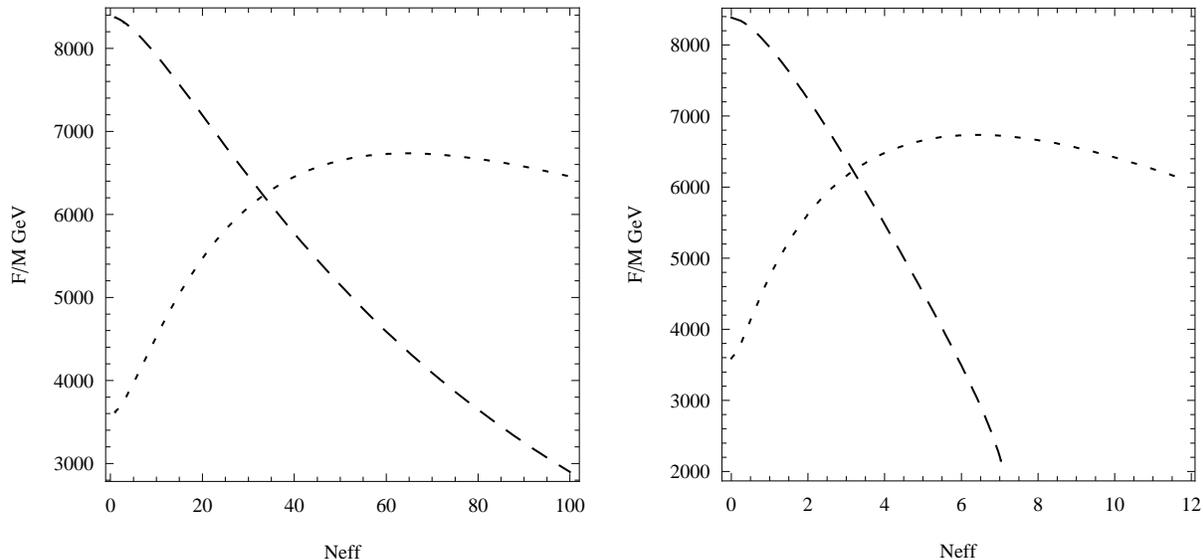}
\caption{\label{10ab} Relevant $D^0-\overline{D^0}$ system bounds on
$F/M$ as a function of $N_{eff}^{(3)}$ from $(\delta^u_{12})_{LL}$
(dashed) and $(\delta^u_{12})_{LL}(\delta^u_{12})_{RR}$ (dotted)
processes, for {\bf 10}-centered models ($M=682 \,{\rm TeV}$). The
left graph refers to Class (a) models and the right graph refers to
Class (b) models.}
\end{center}
\end{figure}

The strongest constraints on the heavy scale are now typically set
by processes involving $(\delta_{12}^u)_{LL}$ terms in
$D^0-\overline{ D^0}$ box diagrams. At relatively high
$N_{eff}^{(3)}$, processes involving
$(\delta_{12}^u)_{LL}(\delta_{12}^u)_{RR}$ become dominant. Plots of
the bounds on $F/M$ as a function of $N_{eff}^{(3)}$ are given in
Figure \ref{10ab}.

Representative bounds can be taken to be
\beq\label{eq:FoverM10}
\begin{split}
&{\rm Class\ (a):}\ \ \ \frac{F}{M}\gsim\ 7\ {\rm TeV}\ \ \ (N^{(3)}_{eff}=28),\\
&{\rm Class\ (b):}\ \ \ \frac{F}{M}\gsim\ 7\ {\rm TeV}\ \ \
(N^{(3)}_{eff}=5),
\end{split}
\eeq
and the bound on the effective scale of SUSY breaking for models of Class (b) is then
\beq \sqrt F\gsim 70\ {\rm TeV}. \eeq
Two-loop contributions then impose the following bounds on $N\equiv
N_{eff}^{(A)}$ at, say, the point in parameter space where all order one
numbers in \eqref{eq:m121c} are equal to one:
\be
\begin{split}\label{eq:N10}
{\rm Class\ (a)}:\ \ \ & N\gsim 28
  \ \ \textrm{for} \  \ \Lambda_S \sim F/M=7\ {\rm TeV},\ \mu_S \sim M=682\ {\rm TeV},\\
{\rm Class\ (b)}:\ \ \ & N\gsim 1 \ \ \textrm{for} \  \ \Lambda_S
\sim \mu_S\sim \sqrt{F}=70\ {\rm TeV}\ (F/M=7\ {\rm TeV}).
\end{split}
\ee At this point in parameter space, Class (a) bounds come
typically from left-handed sleptons, whereas in Class (b) all light
sparticles (except right-handed charged sleptons, which do not
become tachyonic) give comparable bounds.

\subsection{Spectra}\label{ssec:pheno}

We now present some sample spectra for the various partially coupled
models discussed above.  The results are obtained using the program
{\tt SuSpect} \cite{Djouadi:2002ze}, sampling the parameter space
using various effective messenger numbers, specific choices of order
one coefficients coming from the unknown correlation functions in
the CFT sector, and various values of $\tan\beta$. For simplicity,
all CFT sector $N_{eff}^{(A)}$ factors are taken to be equal to $N$
and we always set the scale $M$ to its lower bound. In Class (a)
models, at the classical level, the scale $M$ can be taken to be
much higher than its lower bound, as long as $F/M$ is kept fixed,
but this will modify the RG effects \eqref{eq:RGEmass}.
Additionally, we set the gauge-mediated D-term to zero, since the
qualitative behavior of the spectra is not modified.

\begin{figure}[t]
\begin{center}
\includegraphics[scale=0.6]{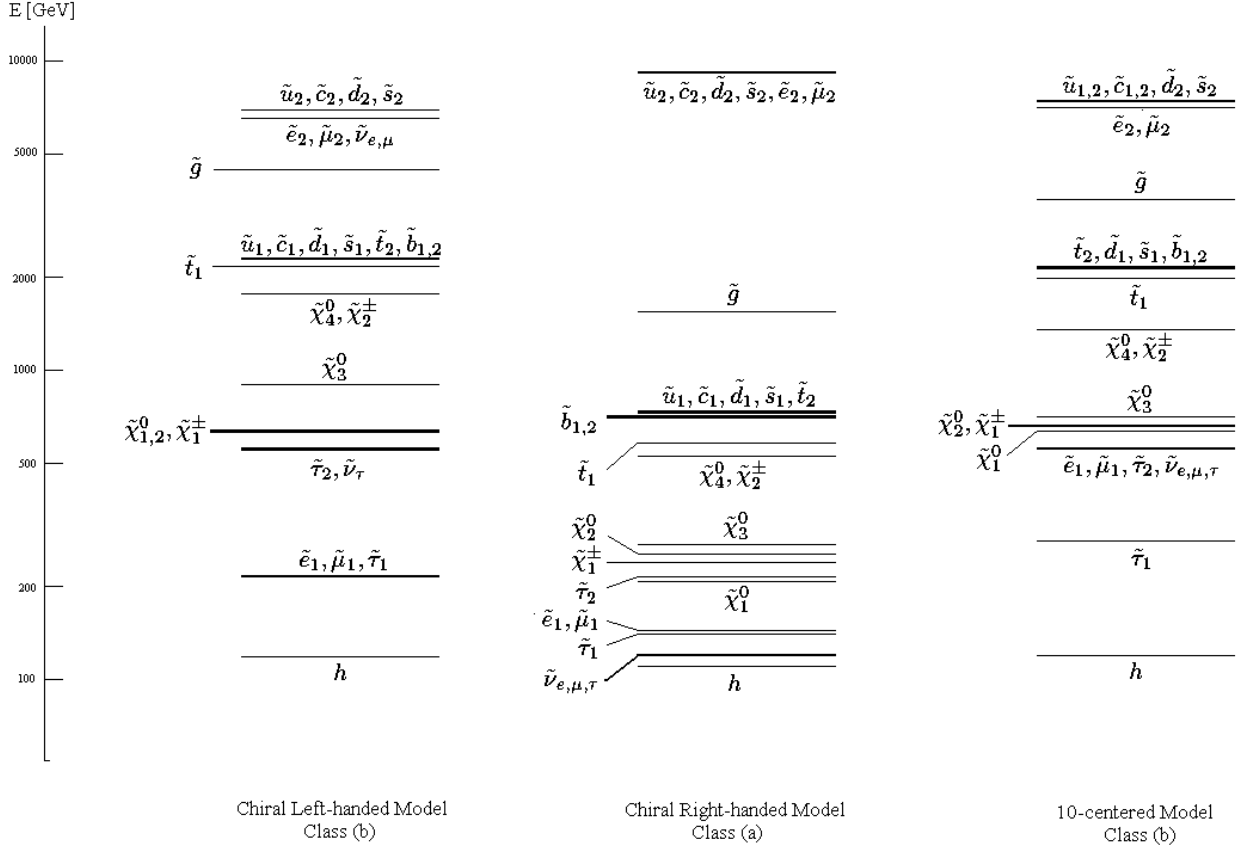}
\caption{\label{fig:spec} The full sparticle spectrum for the models
described in Tables \ref{tab:LHb}, \ref{tab:RHa} and \ref{tab:10b},
for $\tan\beta=10$. }
\end{center}
\end{figure}

In the simulations presented in Tables \ref{tab:LHb}, \ref{tab:RHa},
\ref{tab:RHb}, \ref{tab:10b} and \ref{tab:RHaN} at the end of the
paper, the ${\cal O}(1)$ coefficients coming from the CFT sector in
direct-mediated soft terms (for instance, multiplying the first two
generation sfermion masses-squared in \eqref{eq:m121c}) have all
been taken to be one for simplicity. This implies that all the heavy
sparticles are degenerate, which is certainly not the case
generically in our models. (Models with separable CFT sectors
involving two different scales of SUSY breaking also seem perfectly
viable, though we do not discuss them here.) Otherwise, the spectra
we present are typical. In Table \ref{tab:summary} we show the
inputs for the simulations and references to the corresponding
tables containing representative sparticles of the low-energy
spectra. In the tables, the mass scale of the heavy sparticles is of
order $F/M$.  For completeness, we also include in Table \ref{tab:summary}
the identity of the
NLSP, its mass and the amount of fine-tuning (defined below) in each
type and class of viable model when $\tan\beta=10$. A few examples
of our full spectra are depicted in Figure \ref{fig:spec}.

\begin{table}[t]
\footnotesize{
\begin{center}
\begin{tabular}{|c|c|c|c|c|c|c|c|c|c|} \hline\hline
\ Model &  Class & $\Lambda_S$ & $F/M$ & $\mu_S$ & $N$ &  Table & NLSP  & $m_{NLSP}$   &  Fine-  \cr
        &        &   [TeV]     & [TeV] & [TeV]   &     &        &       & [GeV] &  tuning  \cr
\hline\hline
Left-handed  &  (b) & 45  & 6.5  & 45  & 10 & \ref{tab:LHb} & $\tilde \tau_1$  & 215 & 99   \cr
\hline
Right-handed  &  (a) & 9  & 9  & 220  & 20 & \ref{tab:RHa} & $\tilde \nu$
& 120 & 16   \cr
\hline
Right-handed  &  (b) & 35  & 5  & 35  & 5 & \ref{tab:RHb} & $\tilde \tau_1$  & 153 & 30   \cr
\hline
10-centered  &  (b) & 70 & 7 & 70 & 5 & \ref{tab:10b} & $\tilde \tau_1$  & 280 & 106   \cr
\hline
Right-handed  &  (a) & 9 & 9 & 220 & 15 & \ref{tab:RHaN} & $\tilde \chi^0$  & 168 & 16   \cr
\hline\hline
\end{tabular}
\end{center}}
\caption{Inputs and some results of simulations of partially coupled models. In the above we include references to the corresponding
tables containing the low-energy spectra. For each simulation, the identity of the NLSP,
its mass and the amount of fine-tuning for $\tan\beta=10$ are presented.}
\label{tab:summary}
\end{table}

All spectra obtained in these partially coupled models exhibit
inverted hierarchies.  The sectors in which the first two generation
sfermions are heavy differ amongst the models depending on the
coupling scenario.  Fields that are not directly coupled to the CFT
sector present a mass pattern similar to models of general gauge
mediation. In our setups, the gluino typically interpolates between
the heavy and light scales.  As in models of gauge mediation, the
LSP is always the gravitino.  The NLSP can vary between sneutrino,
stau, neutralino and chargino identities (see Tables
\ref{tab:LHb} through \ref{tab:RHaN} for samples of this, and
\cite{Carpenter:2008he, Rajaraman:2009ga, Abel:2009ve} for recent
NLSP parameter space discussions in general gauge mediation).
Collider studies involving NLSPs of stau, sneutrino, neutralino and
chargino character can be found in \cite{Giudice:1998bp} (and
references therein) and in \cite{Katz:2009qx,Meade:2009qv,Kribs:2008hq}.

In the simulations, the values of $\mu$ and $B_{\mu}$ are determined
in order to reproduce the correct pattern of electroweak symmetry
breaking (namely, the correct VEVs for the two Higgs fields). A
rough order of magnitude estimate of the fine-tuning of the weak
scale in these models is then given by $\sim 2 \mu^2/m_Z^2$.  In
chiral left-handed models and {\bf 10}-centered models $\mu$ is
${\cal O}(600-800)\ {\rm GeV}$ and so these models typically present
fine-tuning at the $1\%$ level, as expected from the relatively
large gluino and stop masses. Chiral right-handed couplings have
lower values of $\mu$, ${\cal O}(300)\ {\rm GeV}$, and so these
models can present fine-tunings milder than $1\%$. A more accurate
quantification of the fine-tuning with respect to a parameter
$\lambda_i$ can be given by the Barbieri-Giudice parameter
$\Delta(m_Z^2; \lambda_i)$ \cite{Barbieri:1987fn}.  In this
language, $\Delta(m_Z^2; \lambda_i)\lsim 100$ corresponds to the
appropriate fine tuning being milder than the percent level.  In the
tables, $\Delta$ denotes the strongest fine-tuning between $\mu^2$
and $B_\mu$, calculated by {\tt SuSpect} \cite{Djouadi:2002ze},
which in all cases considered is $\Delta(m_Z^2; \mu^2)$.
Fine-tunings with respect to other parameters, {\it e.g.} the scale
$\Lambda_S$, are expected in our models to be at most comparable to
the ones presented \cite{Agashe:1997kn}.

Note that in the spectra in Table \ref{tab:summary} left-handed and
{\bf 10}-centered Class (a) models are not presented.  This is
because these models require large values of $N_{eff}^{(A)}$, and so
run into Landau poles. In fact, many (if not all) of our models
exhibit Landau poles for the standard model gauge couplings below
the GUT scale, and many of the models do not exhibit gauge coupling
unification.  We view our models as effective theories valid below
the scale $M_>$, so we do not worry about UV completions above this
scale (except for needing to suppress baryon-violating operators by
a higher scale $M_{pl}$). Models without large numbers of degrees of
freedom in the CFT sector charged under the standard model group can
be safe from Landau poles within the conformal window.  However, our
models that do have large numbers of degrees of freedom in the CFT
sector charged under the standard model group, say $N_{eff}^{(A)}
\gsim 20$, could have Landau poles already below the scale $M_>$,
and then these models are not really valid as effective field
theories (at least not as analyzed above). This means
that some of our Class (a) models are not really self-consistent.

To derive a rough estimate of the upper bound on $N_{eff}^{(A)}$, we
impose that the conformal window is such that a $10^{-5}$ hierarchy
in the up sector can be generated, and demand that the strong
coupling $\alpha_3$ does not blow up in this window.  Combining
this, we find\footnote{We use here equations for the beta functions
that ignore the anomalous dimensions; we know that some of our
fields always have positive anomalous dimensions and others
negative, and we expect some overall correction
 coming from this issue, but we ignore it here since our bounds are up to order one numbers anyway.} the order of
 magnitude constraint $-3+\frac{1}{2}N_{\rm add}\lsim7 \gamma$, where $N_{\rm add}$ stands for additional non-MSSM degrees of
 freedom charged under $SU(3)$ in the conformal window, and $\gamma$ is the sum of the relevant anomalous dimensions in the up
 sector, $\gamma=\frac{1}{2}\left(\gamma_{Q}+\gamma_{\bar{u}}\right)$.  Under the definition of $N_{eff}^{(A)}$ through
 the gaugino masses \eqref{eq:GGMmasses} (relating it in
 weakly coupled theories to, say, the number of pairs of superfields in the fundamental and anti-fundamental
 representations), we obtain for Class (a) models $N_{eff,({\rm a})}^{(3)}\approx \frac{1}{2}(N_{\rm add}+8)$ for fully coupled
 models (where the second term here stems from the fact
 that our definition of $N_{eff, ({\rm a})}^{(A)}$ contains the first two generations of the MSSM, directly
 coupled to the CFT), yielding $N_{eff,({\rm a})}^{(3)}\lsim 7\gamma +7$.  Similarly, in chiral models this gives
$N_{eff,({\rm a})}^{(3)}\lsim 7\gamma+5$, and in {\bf 10}-centered models
$N_{eff,({\rm a})}^{(3)}\lsim 7\gamma+6$. In Class (b) models, the
relation between $N_{\rm add}$ and $N_{eff}^{(A)}$ is less direct,
since $N_{eff}^{(A)}$ only contains the fields that survive to the
lower scale $\sqrt{F}$; clearly $N_{eff,({\rm b})}^{(A)}<\frac{1}{2}N_{\rm
add}$. In all our Class (b) models $N_{eff,({\rm b})}^{(A)}$ is small and
consistent with this, and it seems that it should always be possible
to add few enough degrees of freedom at the higher scale to be
compatible with the bound on $N_{\rm add}$ above.  In the above
simulations we have presented only models that can be consistent
with these bounds.  This issue seems to favor Class (b) models,
although some Class (a) models can also be consistent.


Note that our superconformal sector has an (accidental) $U(1)_R$
symmetry that is broken at the exit from the conformal window. If
this breaking is spontaneous, one may worry that we would have a
light R-axion \cite{Bagger:1994hh} as the corresponding
Nambu-Goldstone boson. However, the R-symmetry in our models is
violated by Yukawa couplings and by terms like ${\cal O}_2 \Phi_1$,
and despite the small coefficient of these terms, this is enough to
raise the R-axion mass to a level where it does not pose any
problems for phenomenology or for cosmology. In any case, we assume
that the R-symmetry is broken at the exit by a large amount, so that
it does not affect the spectrum of soft masses; it may be
interesting to also investigate models where this breaking is small.

\section{Discussion and conclusions}\label{sec:conc}

The Nelson-Strassler mechanism is an elegant means of generating the
pattern of the Yukawa couplings ex-nihilio -- a sector with a
strongly coupled fixed point with a finite basin of attraction
naturally suppresses the Yukawa couplings via RG effects. In this
paper we explored the extension of the Nelson-Strassler mechanism to
a CFT sector which not only determines the Yukawa couplings, but
also triggers SUSY breaking in the MSSM. A coarse look suggests that
``here be dragons'' -- using this CFT sector to set the scale of
SUSY breaking implies that its scale cannot be very high, and then,
since the CFT sector is flavor-dependent (as much as conceivably
possible), the Nelson-Strassler mechanism can run into problems with
FCNC constraints if precautions are not taken when SUSY is broken.

In this work we analyzed these models in detail, focusing on the
constraints from FCNCs, as well as on other problems that are
typical in inverted hierarchy models. An inverted hierarchy is
inevitable in our case, where the strongly coupled SUSY breaking CFT
sector couples directly to (some of) the first two generation
fields, giving rise to large first two generations sfermion masses,
whereas the third generation masses arise from gauge mediation. We
show that despite the apparent difficulties one can construct models
with a viable level of fine-tuning.

In order to achieve low fine-tuning, we had to make several assumptions about the CFT sector:
\begin{enumerate}
\item The CFT sector conserves baryon number.
\item The standard model $U(1)_Y$ is embedded in a non-Abelian group which is a subgroup of the global symmetry group of the CFT sector dynamics. This
is certainly true in GUT models (such as $SU(5)$ models, on which our {\bf 10}-centered model is based), but it can be implemented in other cases as well.
\item In order to suppress flavor-violating effects, including those coming from mixings
at the exit from the conformal window (accomplishing a graceful exit), we require two features of
the CFT sector,
both compatible with the Nelson-Strassler construction:
\begin{enumerate}
\item The CFT sector is separable (for instance by coupling each of the first two generations to a separate CFT sector), meaning that the CFT sector does not directly produce any flavor changing in the interaction basis.
\item The models are partially coupled -- only a subset of the fields in the first two generations couple directly to the CFT sector.
\end{enumerate}

\item The CFT sector has two scales, namely the scale of SUSY breaking $\sqrt{F}$ is much smaller than the scale $M$ where the first two generations decouple from the CFT sector.
\end{enumerate}

In this paper we did not attempt to construct a model that obeys all
of these requirements (as well as a Nelson-Strassler mechanism and
dynamical SUSY breaking). It seems that such a model should be quite
complicated, but we do not see any obstruction in principle to the
construction of such models (see \cite{Ibe:2005pj,Schmaltz:2006qs}
for related, though {\it a priori} not directly applicable, examples
in this direction). Our discussion assumed a strongly coupled CFT
sector, but otherwise it is completely general; in particular it
applies to models \cite{DualNSModels} where this sector has a weakly
coupled description in higher dimensions. In Class (b) models it is
possible that the CFT sector below the scale $M$, and in particular
at the scale of SUSY breaking, could become weakly coupled, but we
do not discuss this possibility here. As discussed above, our models
generally exhibit Landau poles below the GUT scale, but suitable UV
completions can perhaps be found, {\it e.g.} via Seiberg duality.

We believe that even without building explicit models, our
constructions should have various distinctive features that could
perhaps be tested at the LHC. It would be interesting to perform a
general analysis of the phenomenology of these models, but this is
beyond the scope of the present paper. Let us just mention here the
general form of the spectrum that the assumptions above imply. We
always have a partial inverted hierarchy for the sparticles, where
the masses of the scalar components of the superfields coupled
directly to the CFT sector are around $10$ TeV. The gluinos are
lighter, with a mass of a few TeV, and the rest of the superpartners
are all around $2$ TeV or below. The NLSP can be a stau,
sneutrino, neutralino or chargino, as in general gauge mediation.  In
comparison to gauge mediation, our spectra differ by the heaviness
of some of the first two generation fields. The partial inverted
hierarchy that we presented here could be even more partial in
separable models with only one SUSY-breaking component, but we do
not analyze this case here.

In comparison to other models with an inverted hierarchy, here the
light sparticles have a mass spectrum dictated specifically by gauge
mediation.  Additionally, in our partially-coupled models only some
chiralities exhibit an inverted hierarchy (see
\cite{Ambrosanio:1997ky} for an example of a previous model with
related characteristics). We also have rather heavy gluinos.
Recently, a suggestion for joining a different model of flavor
physics with supersymmetry breaking \cite{Compositemodels} was
realized in \cite{Franco:2009wf,Craig:2009hf}; there are many
similarities of these models to what we find here, but also some
differences. Note in particular that in our models the difference
between the three generations of the standard model is generated
purely dynamically, while in the models of
\cite{Compositemodels,Franco:2009wf,Craig:2009hf} one must put in by
hand that only the third generation fields are elementary at high
energies. Recent LHC-oriented phenomenological studies of inverted
hierarchy models have appeared in \cite{Desai:2009ex}.

In this paper we have thus far ignored the bounds on new physics coming from
CP-violating processes. Even if the hidden sector conserves CP,
this is not really justified, since the order one CP-violating phase in
the CKM matrix generically means that the CP-violating flavor-changing
processes in our model will be of the same order as the CP-conserving
ones.\footnote{We thank P. Paradisi and Y. Nir for discussions on this issue.} Thus, in
the general case the bounds used for CP-conserving processes would be replaced by the bounds including CP-violating processes.  The fine-tuning required in our models would then be enhanced by a factor of order 25, depending on the precise model, leading to unacceptably large fine-tuning.  The level of fine-tuning can be reduced (with or without the CP-violating operators) by making some additional assumptions about the structure of our models.  For instance, one could assume that at high energies there is an
$SU(3)\times SU(3)$ global symmetry that guarantees that the
up-Yukawa couplings (or the down-Yukawa couplings) are proportional
to the identity matrix (this symmetry is then broken by the couplings
to the hidden sector). Or, one could assume that there is some sort of
alignment that guarantees that the up-Yukawa couplings (or the
down-Yukawa couplings) are diagonal in the same basis as the couplings to
the hidden sector (this can naturally arise in extra dimensional scenarios,
where different fields are located at different positions in the extra
dimensions). In both cases, most of the discussion in our paper is not
modified, but flavor-changing processes involving either up or down quarks are suppressed. Since in most of our models one of the two types of processes gives much stronger bounds than the other, this allows us to reduce the amount of fine-tuning by a factor of 8 or so.  Note that in all cases the bounds coming from flavor-conserving CP-violating processes, such as electric dipole moments \cite{Chang:1998uc,Ellis:2008zy} (see also {\it e.g.} \cite{flavoredEDM}), are less constraining or at most comparable. We leave a detailed discussion of the implications of these additional assumptions, and of the construction of such models, to future work.

Finally, we should emphasize that in this paper we did not attempt
to address the $\mu/B_\mu$ problem, which is generic in models of
gauge mediation. Presumably, recent solutions to this problem (such
as \cite{Murayama:2007ge}) can be applied to our models as well.
Note that in the natural scenario in which the Higgs fields do not
couple directly to the CFT sector, $B_{\mu}=0$ at leading order at
the scale $\mu_S$, and it was recently claimed \cite{Abel:2009ve}
that such a scenario is not impossible if $\tan\beta$ is large
enough. We leave a detailed analysis of the Higgs sector and its
couplings to future work.

\pagebreak

\centerline{}
\centerline{\bf Acknowledgements}

We would like to thank Oram Gedalia, Vadim Kaplunovsky, Lorenzo Mannelli, Gilad Perez and Tomer
Volansky for many useful conversations, and Shamit Kachru, Zohar
Komargodski and Yossi Nir for fruitful discussions and comments on a draft of this
paper. This work was supported in part by the Israel-U.S. Binational
Science Foundation, by a center of excellence supported by the
Israel Science Foundation (grant number 1468/06), by a grant (DIP
H52) of the German Israel Project Cooperation, and by the Minerva foundation with funding from the Federal German Ministry for Education and Research.

\vskip 50pt



\begin{table}[h]
\footnotesize{
\begin{center}
\begin{tabular}{|c||c|c|c|c|c|c|c|c|c|c|c|c|c|c|c|c|} \hline\hline
\ $\tan \beta$ &  $m_h$ & $m_{\tilde\chi^0_1}$ &
$m_{\tilde\chi^+_1}$ & $m_{\tilde t_1}$ & $m_{\tilde b_1}$ &
$m_{\tilde \tau_1}$ & $m_{\tilde u_1}$ & $m_{\tilde d_1}$ &
$m_{\tilde e_1}$ & $M_3$ & $\mu$ & $\Delta$  \cr \hline\hline
 3 &  108 & 754 & 766 & 2150 &  2274 & 215 & 2281 & 2275 & 216 & 4446 & 761 & 218   \cr \hline
10  &  119 & 631 & 636 & 2165 & 2272 & 215 & 2281 & 2275 & 217 &
4445 & 629 & 99 \cr \hline\hline
\end{tabular}
\end{center}}
\caption{Example of a spectrum for Class (b) models with $N=10$,
when coupling only left-handed first two generation fields to a
decomposable CFT sector.  We present representative sparticles of
the low energy spectrum. The appropriate heavy sparticle masses are of
order $F/M$.  The results are obtained using $F/M=6.5$ TeV,
$\mu_S=\Lambda_S=45$ TeV for two different values of $\tan\beta$,
choosing all unknown coefficients coming from the CFT sector to
equal one. The entries for $\tilde u_1, \tilde d_1, \tilde e_1$
refer to $\tilde c_1, \tilde s_1, \tilde \mu_1$ as well.  $\Delta$
is the fine-tuning of $m_Z^2$ with respect to $\mu^2$. All
dimensionful quantities are given in GeV.  } \label{tab:LHb}
\end{table}

\begin{table}[h]
\footnotesize{
\begin{center}
\begin{tabular}{|c||c|c|c|c|c|c|c|c|c|c|c|c|c|c|c|c|} \hline\hline
\rule{0pt}{1.2em}%
\ $\tan \beta$ &  $m_h$ & $m_{\tilde\chi^0_1}$ &
$m_{\tilde\chi^+_1}$ & $m_{\tilde t_1}$ & $m_{\tilde b_1}$ &
$m_{\tilde \tau_1}$ & $m_{\tilde u_1}$ & $m_{\tilde d_1}$ &
$m_{\tilde e_1}$ & $m_{\tilde \nu}$  & $M_3$ & $\mu$ & $\Delta$  \cr
\hline\hline 3  &  98 & 218 & 277 & 567 & 697 & 141 & 722 & 726 &
142 & 123 & 1533 & 295 & 34   \cr \hline 10  &  111 & 206 & 238 &
580 & 696 & 140 & 722 & 727 & 143 & 120 & 1533 & 246 & 16   \cr
\hline\hline
\end{tabular}
\end{center}}
\caption{The same for Class (a) models with $N=20$, when coupling
only the right-handed first two generation fields to a decomposable
CFT sector, using $F/M=9$~TeV, $\mu_S=220$~TeV and
$\Lambda_S=9$~TeV. }
 \label{tab:RHa}
\end{table}

\begin{table}[h]
 \footnotesize{
\begin{center}
\begin{tabular}{|c||c|c|c|c|c|c|c|c|c|c|c|c|c|c|c|} \hline\hline
\ $\tan \beta$ &  $m_h$ & $m_{\tilde\chi^0_1}$ &
$m_{\tilde\chi^+_1}$ & $m_{\tilde t_1}$ & $m_{\tilde b_1}$  &
$m_{\tilde \tau_1}$ & $m_{\tilde u_1}$ & $m_{\tilde d_1}$ &
$m_{\tilde e_1}$ & $m_{\tilde \nu}$ & $M_3$ & $\mu$ & $\Delta$  \cr
\hline\hline
 3 &  101 & 314 & 394 & 1018  & 1099 & 154 & 1129 & 1131 & 277 & 268 & 1860 & 408 & 64   \cr \hline
10 & 114 & 300 & 336 & 1029 & 1097 & 153 & 1129 & 1132 & 277 & 266 &
1860 & 342 & 30 \cr \hline\hline
\end{tabular}
\end{center}}
\caption{The same for Class (b) models with $N=5$, when coupling
only the right-handed first two generations to a decomposable CFT
sector, using $F/M=5$ TeV and $\mu_S=\Lambda_S=35$~TeV. }
 \label{tab:RHb}
\end{table}

\begin{table}[h]
\footnotesize{
\begin{center}
\begin{tabular}{|c||c|c|c|c|c|c|c|c|c|c|c|c|c|c|} \hline\hline
\ $\tan \beta$ &  $m_h$ & $m_{\tilde\chi^0_1}$ &
$m_{\tilde\chi^+_1}$ & $m_{\tilde t_1}$  & $m_{\tilde b_1}$ &
$m_{\tilde \tau_1}$ 
&$m_{\tilde d_1}$ &
$m_{\tilde e_1}$  & $m_{\tilde\nu}$ & $M_3$ & $\mu$ & $\Delta$  \cr
\hline\hline 3  &  107 & 664 & 783 & 1943 & 2112 & 281 
&
2113 & 558 & 553 & 3538 & 782 & 231   \cr \hline 10  &  119 & 629 &
655 & 1959 & 2110 & 280 
& 2113 & 558 & 552 & 3538 & 650 &
106   \cr \hline\hline
\end{tabular}
\end{center}}
\caption{The same for Class (b) models with $N=5$ in the {\bf
10}-centered coupling scenario with a decomposable CFT sector, using
$F/M=7$~TeV and $\mu_S=\Lambda_S=70$~TeV. } \label{tab:10b}
\end{table}

\begin{table}[h]
\footnotesize{
\begin{center}
\begin{tabular}{|c||c|c|c|c|c|c|c|c|c|c|c|c|c|c|c|} \hline\hline
\rule{0pt}{1.2em}%
\ $\tan \beta$ &  $m_h$ & $m_{\tilde\chi^0_1}$ &
$m_{\tilde\chi^+_1}$ & $m_{\tilde t_1}$  & $m_{\tilde b_1}$ &
$m_{\tilde \tau_1}$ & $m_{\tilde u_1}$ & $m_{\tilde d_1}$ &
$m_{\tilde e_1}$ & $m_{\tilde \nu}$  & $M_3$ & $\mu$ & $\Delta$  \cr
\hline\hline
3  &  97 & 170 & 267 & 522 & 666 & 190 & 702  & 706 & 191 & 178 & 1181 & 307 & 37   \cr
\hline 10  &  110 & 168 & 234 & 538 & 664 & 188 & 703 & 707 & 192 & 175 & 1181 & 252 & 16   \cr
\hline\hline
\end{tabular}
\end{center}}
\caption{Example of a spectrum for Class (a) models with $N=15$, in
the chiral right-handed coupling scenario with a decomposable CFT
sector. The results are obtained using $F/M=9$~TeV, $\mu_S=220$~TeV
and $\Lambda_S=9$~TeV, choosing the ${\cal O}(1)$ numbers equal to
$1$ for direct-mediated soft terms and $3$ for gauge-mediated ones.}
\label{tab:RHaN}
\end{table}


\clearpage


\end{document}